\documentclass[aps,floats,nofootinbib,amssymb,preprint,superscriptaddress]{revtex4}

\usepackage{graphicx}

\newcommand{\be}{\begin{equation}}
\newcommand{\ee}{\end{equation}}
\newcommand{\bea}{\begin{eqnarray}}
\newcommand{\eea}{\end{eqnarray}}

\renewcommand{\check}{({\bf This statement needs to be checked!!})}

\newcommand{\ba}{\begin{array}}
\newcommand{\ea}{\end{array}}
\newcommand{\bi}{\begin{itemize}}
\newcommand{\ei}{\end{itemize}}
\newcommand{\ben}{\begin{enumerate}}
\newcommand{\een}{\end{enumerate}}

\preprint{
\hbox to \hsize{
\hfill$\vcenter{\hbox{\bf MADPH-08-1515}
\hbox{\bf ANL-HEP-PR-09-3}
\hbox{\bf NUHEP-TH/09-02}
              \hbox{February 2009}}$}
}

\begin{document}
\title{\vspace*{.5in}Higgs ID at the LHC}

\author{Vernon~Barger}
\email{barger@pheno.physics.wisc.edu}
\affiliation{Department of Physics, University of Wisconsin, 1150 University
Avenue, Madison, Wisconsin 53706 USA}

\author{Heather~E.~Logan}
\email{logan@physics.carleton.ca}
\affiliation{Ottawa-Carleton Institute for Physics, Carleton University,
Ottawa K1S 5B6 Canada}

\author{Gabe~Shaughnessy}
\email{g-shaughnessy@northwestern.edu}
\affiliation{Department of Physics, University of Wisconsin, 1150 University
Avenue, Madison, Wisconsin 53706 USA}
\affiliation{Northwestern University, Department of Physics and Astronomy, 2145 Sheridan Road, Evanston, IL 60208 USA}
\affiliation{HEP Division, Argonne National Lab, Argonne IL 60439 USA}

\begin{abstract}

We make a complete catalog of extended Higgs sectors involving SU(2)$_L$ doublets and singlets, subject to natural flavor conservation. In each case we present the couplings of a light neutral CP-even Higgs state $h$ in terms of the model parameters, and identify which models are distinguishable in principle based on this information.  We also give explicit expressions for the model parameters in terms of $h$ couplings and exhibit the behaviors of the couplings in the limit where the deviations from the Standard Model Higgs couplings are small.  Finally we discuss prospects for differentiation of extended Higgs models based on measurements at the LHC and ILC and identify the regions in which these experiments could detect deviations from the SM Higgs predictions.
\end{abstract}
\thispagestyle{empty}
\maketitle

\section{Introduction}

The Standard Model (SM) of particle physics has provided a remarkably
successful description of electroweak data up to present energies.
While the minimal SM implementation of electroweak symmetry breaking
(EWSB) relies on a single SU(2)$_L$ doublet Higgs field, the dynamics
of the EWSB sector have not yet been directly probed and extensions of
the SM allow for a wide variety of extended Higgs sectors consistent
with all existing data.\footnote{Models with extra Higgs doublets
and/or singlets can yield better agreement with electroweak precision
measurements than the SM;
cf.\ Refs~\cite{Bertolini:1985ia,Hollik:1986gg,Heinemeyer:2006px,Profumo:2007wc,Barger:2007im,Fukugita:2008ba}.}
Indeed, models that address the hierarchy problem -- the extreme
instability of the SM Higgs mass parameter to radiative corrections
-- contain additional fundamental or composite
scalar particles at or below the TeV scale.

If a Higgs-like state is discovered, the next priority will be to test
the Higgs mechanism of mass generation by measuring its couplings to
SM particles~\footnote{Determining the nature of a new observed spin-less state that is not a Higgs is also challenging~\cite{Burgess:1999ha}.}.  Experimental sensitivity to the Higgs couplings comes
from measurements of the Higgs production cross sections, decay
branching fractions, and total width.  Prospects for the extraction of
Higgs couplings from experimental data have been studied for the CERN
Large Hadron Collider
(LHC)~\cite{Djouadi:2000gu,Zeppenfeld:2000td,Belyaev:2002ua,Duhrssen:2004cv},
International Linear $e^+e^-$ Collider (ILC)~(for a review and
references see Ref.~\cite{Djouadi:2007ik}), photon
collider~\cite{Monig:2005uj,Niezurawski:2002jx}, and muon
collider~\cite{Barger:2001mi}; for a recent review of Higgs boson
production and decay at these machines see also
Ref.~\cite{Djouadi:2005gi}.  These measurements can be used to test
the consistency of the measured Higgs couplings with SM predictions,
and, if a discrepancy is found, to constrain the possible nature of the
extended model.  Different models give rise to different patterns of
Higgs coupling deviations; in other words, they occupy different
``footprints'' in the space of measurable Higgs couplings.  By
identifying the footprint of each different extended Higgs sector in
the space of Higgs couplings, we can determine whether the models can
be distinguished in principle and establish a framework to map back
any observed pattern of Higgs coupling deviations onto the appropriate
underlying model.

Our aim is to make a complete catalog of extended Higgs sectors
based on the patterns of coupling shifts of a single neutral Higgs
state $h$.  In this paper we limit ourselves to models that contain
only Higgs doublets and/or singlets.  This allows us to avoid
tree-level violation of custodial SU(2) symmetry and the resulting
stringent constraints from the $\rho$ parameter that arise, e.g., in
models containing Higgs triplets~\cite{Chen:2005jx}.  We also limit
our study to models that obey the Glashow-Weinberg-Paschos
condition~\cite{Glashow:1976nt,Paschos:1976ay} for natural flavor
conservation, which requires that all fermions of the same electric charge get their
mass from exactly one Higgs doublet.  This allows us to avoid tree-level flavor-changing neutral
Higgs interactions and the resulting stringent constraints from
low-energy flavor physics.  We also neglect the possibility of CP
violation~\cite{Accomando:2006ga}, and assume that the state $h$ is pure CP-even.  Of course,
many viable extended Higgs models exist outside these categories;
their inclusion into the framework of Higgs coupling footprints
presented here would make an obvious future extension of this work.

Within the above constraints we enumerate the complete set of models that
can arise and present formulas for the shifts in the couplings of $h$
to SM particles relative to their SM values.  We consider only what
can be learned from one Higgs state, $h$; of course, observation of
additional Higgs states (CP-even, CP-odd, or charged) or other new
particles will complement this information.  We focus on the couplings that arise from dimension-four operators, in particular the couplings of $h$ to $W$ or $Z$ boson pairs and to fermion pairs.  Note that the $hWW$ and $hZZ$ couplings are modified from their SM values by a \emph{common} multiplicative factor in any model containing only Higgs doublets and singlets.  Similarly, our assumption of natural flavor conservation implies that the $h \bar u_i u_i$ couplings for the three generations of up-type quarks are modified by a \emph{common} multiplicative factor; the same holds for down-type quarks and for charged leptons.  We do not give explicit results for the loop-induced Higgs couplings to gluon or photon pairs or $\gamma Z$ because new physics at higher scales can generate additional effective couplings to these final states comparable in strength to those induced by SM loops~\cite{Manohar:2006gz}\footnote{Indeed, efforts to distinguish new physics running in loops involved in production and decay of a lone Higgs state have been made for selected models~\cite{Hsieh:2008jg}.}.  These loop-induced couplings are nevertheless important because they provide experimental access to the relative signs of the dimension-four couplings.  Similarly, higher scale physics can generate dimension-six $hWW$ and $hZZ$ operators~\cite{Hagiwara:2000tk,Barger:2003rs,Dutta:2008bh,Kanemura:2008ub,Qi:2008ex}; we do not include these effects in our analysis.

This paper is organized as follows.  In the next section we briefly
review the Higgs couplings in the SM and introduce the notation and
general framework that will be used to describe the extended models.
We then proceed to the discussion of the extended models in
Secs.~\ref{sect:1hdm}, \ref{sect:2hdm} and \ref{sect:3hdm}.  For each
model we present the couplings of $h$ in terms of its composition and
the vacuum expectation values (vevs) of the doublets in the model;
where possible we also invert these relations to find explicit
expressions for the model parameters in terms of the Higgs couplings.
We identify the coupling patterns that allow the different models to
be distinguished and specify the sets of models that cannot be
distinguished based on the couplings of only one state. We also give
expansions for the couplings near the decoupling
limit~\cite{Barger:1992zy,Haber:1994mt} in which the deviations of the
couplings from their SM values are small.  In Sec.~\ref{sect:radcorr}, we discuss the implications radiative corrections would have on our results.

We finish in Sec.~\ref{sect:disc} by comparing the predictions of the
individual models to each other and to the expected experimental
sensitivity of the LHC and ILC.  We present a decision tree for
identifying the underlying model based on the couplings of $h$ and
point out which models cannot be distinguished even if the couplings
of $h$ were known exactly.  Overall we find 15 models (or sets of
models) with extended Higgs sectors that are distinguishable in
principle in at least part of their parameter spaces.  We also discuss
the prospects for model differentiation based on the expected accuracy
of Higgs coupling measurements at the LHC and ILC.  For representative
models, we plot the regions in which $h$ would appear SM-like given
the expected experimental uncertainties at the LHC
(Fig.~\ref{fig:hmodels-decoup}) and ILC
(Fig.~\ref{fig:hmodels-decoup-ilc}). We also provide a summary table
showing the decoupling behavior of the $h$ partial widths. We end with
a brief summary of our conclusions.

We find it most straightforward to organize the models on the basis of
the structure of the Yukawa Lagrangian -- in particular, the number of
different Higgs doublet(s) involved in fermion mass generation.
In Sec.~\ref{sect:1hdm} we consider models in which the fermion masses
are generated by the vev of only one Higgs
doublet.  Models with this characteristic are:
\begin{itemize}
\item The SM, in which the Higgs sector consists of only one SU(2)$_L$
doublet that gives masses to the $W$ and $Z$ bosons and all the quarks
and charged leptons.  This is the simplest realization of the Higgs
mechanism.

\item The SM extended with one or more singlet
scalars~\cite{Silveira:1985rk,Davoudiasl:2004be,BahatTreidel:2006kx,OConnell:2006wi,Barger:2007im}.
This model yields an overall reduction in Higgs couplings due to doublet-singlet mixing and its phenomenology has been studied extensively.  Models
with similar light Higgs phenomenology include unparticle models in
which Higgs-unparticle mixing can suppress
couplings~\cite{Delgado:2008px} and Randall-Sundrum models in which
Higgs-radion mixing also leads to reduced
couplings~\cite{Csaki:2000zn,Kribs:2001ic,Csaki:2007ns}.

\item The Type-I two Higgs doublet model (2HDM), in which only one
doublet couples to fermions, while both doublets are involved in the
generation of the $W$ and $Z$ boson
masses~\cite{Georgi:1978wr,Haber:1978jt}.  The sharing of the vev
between two doublets can have a dramatic effect on the coupling
pattern of the Higgs boson.

\item A Type-I 2HDM extended with one or more singlet scalars.

\item A Type-I 2HDM extended with one or more additional doublets.
\end{itemize}

In Sec.~\ref{sect:2hdm} we consider models in which the fermion masses
are generated by the vevs of two different Higgs doublets.  Models with
this characteristic are:
\begin{itemize}
\item The Type-II 2HDM, in which one doublet generates the masses of
the up-type quarks while a second doublet generates the masses of the
down-type quarks and charged
leptons~\cite{Lee:1973iz,Fayet:1974fj,Peccei:1977hh,Fayet:1976cr,Gunion:1989we,Carena:2002es}.
This fermion coupling structure appears at tree level in the minimal
supersymmetric standard model (MSSM), contributing to the great
popularity of the Type-II 2HDM.  In the MSSM, radiative corrections
involving supersymmetric particles can induce potentially significant
couplings of the bottom quarks to the ``wrong'' Higgs doublet~\cite{Hall:1993gn,Hempfling:1993kv}; while
this feature formally puts the MSSM Higgs sector outside our
requirement of natural flavor conservation, we nevertheless consider
the features of this extension as well.

\item A Type-II 2HDM extended with one or more singlet
scalars~\cite{Gunion:1989we,Das:1995df,Kiers:1998ry,Wu:1999nc,Ellwanger:2004xm,Schabinger:2005ei,Barger:2006dh,Xiao:2006dq}.\footnote{Singlet
extensions of the Higgs sector are also popular in supersymmetric
models~\cite{Haber:1984rc,Drees:1988fc,Martin:1997ns,Drees:2004jm,Binetruy:2006ad,Baer:2006rs,Barger:2006dh,Barger:2006sk}.}

\item A Type-II 2HDM extended with one or more additional
doublets~\cite{Drees:1988fc}.

\item The ``flipped'' and ``lepton-specific'' 
2HDMs~\cite{Barnett:1983mm,Barnett:1984zy,Grossman:1994jb}, 
in which the
coupling assignments of the two Higgs doublets to up-type quarks,
down-type quarks, and charged leptons are varied relative to the usual
Type-II 2HDM.  In the flipped 2HDM, one doublet generates the masses
of the up-type quarks and charged leptons while the second doublet
generates the masses of the down-type quarks.  In the lepton-specific
2HDM, one doublet generates the masses of both up-type and down-type
quarks while the second doublet generates the masses of the leptons.
We also consider extensions of these two models with additional
doublets and/or singlet scalars.
\end{itemize}

In Sec.~\ref{sect:3hdm} we consider models in which the fermion masses
are generated by the vevs of three different Higgs doublets.  Models
with this characteristic are:
\begin{itemize}
\item A ``democratic'' three Higgs doublet model (3HDM-D), in which one
doublet generates the masses of up-type quarks, a second doublet generates
the masses of down-type quarks, and a third doublet generates the masses
of the charged leptons.

\item The 3HDM-D extended with one or more singlet scalars.

\item The 3HDM-D extended with one or more additional doublets.
\end{itemize}

The physical Higgs boson content of these models can be summarized as
follows.  Consider a model that contains $N_d$ complex doublets and
$N_s$ singlets, $N_c \leq N_s$ of which are complex.  After removing
the unphysical charged and neutral Goldstone bosons, this model
contains $N_d - 1$ charged Higgs bosons $H^{\pm}_i$, $N_d + N_s$
CP-even neutral Higgs bosons $H^0_i$, and $N_d + N_c - 1$ CP-odd
neutral Higgs bosons $A^0_i$.  The state that we consider throughout
is one of the CP-even neutral Higgs bosons, denoted $h$.  Our results
can be applied to any of the CP-even neutral Higgs bosons $H^0_i$.
We make no assumptions about whether additional states can be discovered
at the LHC.


\section{Higgs couplings in the Standard Model and beyond}
\label{sec:framework}

The Higgs doublet of the SM is given by
\begin{equation}
\Phi = \left( \begin{array}{c}
  \phi^+ \\
  (\phi^{0,r} + v_{SM} + i \phi^{0,i})/\sqrt{2}     \end{array} \right),
\end{equation}
where the vev of the Higgs field is $v_{SM} = 246$ GeV.  The couplings
of the Higgs to SM fermions are given by the Yukawa Lagrangian,
\begin{equation}
\mathcal{L}_{Yuk} = - y_e \bar e_R \Phi^{\dagger} L_L
- y_d \bar d_R \Phi^{\dagger} Q_L
- y_u \bar u_R \tilde \Phi^{\dagger} Q_L
+ {\rm h.c.},
\end{equation}
where $Q_L = (u_L, d_L)^T$, $L_L = (\nu_L, e_L)^T$, and $\tilde \Phi$
is the conjugate Higgs multiplet,
\begin{equation}
\tilde \Phi \equiv i \sigma_2 \Phi^*
= \left( \begin{array}{c} (\phi^{0,r} + v_{SM} - i \phi^{0,i})/\sqrt{2} \\
  -\phi^- \end{array} \right).
\end{equation}
These couplings generate the fermion masses, $m_f = y_f v_{SM} /
\sqrt{2}$, and the Feynman rules for the couplings of the physical SM
Higgs boson $h = \phi^{0,r}$ to the fermions,
$-i y_f/\sqrt{2} = -i m_f / v_{SM} \equiv -i g_f^{SM}$.

The couplings of the Higgs to the $W$ and $Z$ bosons arise from the
covariant derivative terms in the Lagrangian,
$\mathcal{L} = \left| \mathcal{D}_\mu \Phi \right|^2$,
where the covariant derivative is given by
\begin{equation}
\mathcal{D}_{\mu} = \partial_{\mu}
- i \frac{g}{\sqrt{2}}\left( W_{\mu}^+ T^+ + W_{\mu}^- T^- \right)
- i \sqrt{g^2 + g^{\prime 2}} Z_{\mu} \left( T^3 - \sin^2 \theta_W Q \right)
- i e A_{\mu} Q.
\end{equation}
This term generates the $W$ and $Z$ boson masses,
\begin{equation}
m_W = \frac{g \, v_{SM}}{2}, \qquad \qquad
m_Z = \frac{ \sqrt{g^2 + g^{\prime 2}} \, v_{SM}}{2},
\end{equation}
and the Feynman rules for the couplings of $h$ to $W$ or $Z$ boson
pairs, given by $i g_V^{SM} g_{\mu\nu}$ where
\begin{equation}
g_W^{SM} = \frac{g^2 v_{SM}}{2}, \qquad \qquad
g_Z^{SM} = \frac{(g^2 + g^{\prime 2}) v_{SM}}{2}.
\end{equation}

We now introduce our general framework for the extended models considered
in this paper.  In a model with multiple Higgs doublets, we can define the neutral, CP-even Higgs mass eigenstate under consideration as
\begin{equation}
h = \sum_i a_i \phi_i,
\label{eq:hsum}
\end{equation}
where $\phi_i \equiv \phi_i^{0,r}$ is the properly normalized
real neutral component of doublet $\Phi_i$.  The coefficients $a_i
\equiv \langle h | \phi_i \rangle$ are constrained by the usual
quantum mechanical requirement that $h$ be properly normalized:
\begin{equation}
\sum_i |a_i|^2 = 1.
\label{eq:asum}
\end{equation}
In such a model, the vev that gives rise to the $W$ and $Z$ masses is
in general shared among the doublets.  We define the ratio of each
doublet's vev to $v_{SM}$ as
\begin{equation}
b_i \equiv \frac{v_i}{v_{SM}}, \qquad \qquad
\sum_i b_i^2 = 1,
\label{eq:bsum}
\end{equation}
where the second condition is required to obtain the correct $W$ and
$Z$ masses.  In the absence of CP violation, which we assume
throughout, the quantities $b_i$ can all be chosen real and positive.
Eq.~\ref{eq:bsum} can also be thought of as a normalization
requirement.  In models containing only doublets, one can define a
linear transformation to a ``Higgs basis''~\cite{Davidson:2005cw} in
which only one doublet, $\Phi_v$, carries a nonzero vev.  The ratios $b_i$
are then given by $b_i = \langle \phi_i | \phi_v \rangle$, such that
$\phi_v = \sum_i b_i \phi_i$ and the condition $\sum_i b_i^2 = 1$
follows from unitarity.

In models containing both doublets and singlets, these relations are modified
as follows.  Because the Higgs state $h$ can contain a singlet admixture,
the sums in Eqs.~\ref{eq:hsum} and \ref{eq:asum} must include the singlet
states as well as the doublets:
\begin{equation}
h = \sum_{\rm doublets, \, singlets} a_i \phi_i, \qquad \qquad
\sum_{\rm doublets, \, singlets} |a_i|^2 = 1,
\label{eq:sasum}
\end{equation}
where now $\phi_i$ can also represent the real neutral component of a
singlet.  
The $W$ and $Z$ masses, however, are generated only by the
vevs of doublets, so that the sum in Eq.~\ref{eq:bsum} is restricted
to run over doublet vevs only:
\begin{equation}
\sum_{\rm doublets \, only} b_i^2 = 1.
\label{eq:sbsum}
\end{equation}
While singlet scalars can have vevs of their own, these singlet vevs play
no role in our analysis and will be ignored.

In such an extended model, the couplings of the real neutral 
states $\phi_i$ to $W$ or $Z$ boson pairs arise from the covariant
derivative terms for the doublets,
\begin{equation}
\mathcal{L} = \sum_{\rm doublets} \left| \mathcal{D}_\mu \Phi_i
\right|^2.
\end{equation}
After the mixing in Eq.~\ref{eq:sasum}, the coupling of the physical
state $h$ to $W$ or $Z$ boson pairs is controlled by the overlap of
$h$ with the doublet $\phi_v$ that carries the vev in the Higgs
basis,
\begin{equation}
  g_V^h = g_V^{SM} \langle h | \phi_v \rangle,
  \label{eq:gV}
\end{equation}
where $V = W$ or $Z$.
Inserting a complete set of states, we obtain,
\begin{equation}
g_V^h = g_V^{SM} \sum_i \langle h | \phi_i \rangle
        \langle \phi_i | \phi_v \rangle
	= g_V^{SM} \sum_{\rm doublets \, only} a_i b_i,
\end{equation}
where the restriction of the sum to run over only the doublets arises
because $\phi_v$ cannot contain a singlet admixture.  Note that
$g_W^h/g_W^{SM} = g_Z^h/g_Z^{SM}$.

We note here that the familiar 2HDM sum rule~\cite{Gunion:1990kf} for
Higgs couplings to gauge bosons can be generalized to models
containing arbitrary numbers of doublets and singlets.  Summing over
all CP-even neutral states and using Eq.~\ref{eq:gV}, we have,
\begin{equation}
  \sum_{H^0_i} (g_V^{H^0_i})^2 
  = (g_V^{SM})^2 \sum_{H^0_i} |\langle H^0_i | \phi_v \rangle |^2 
  = (g_V^{SM})^2,
  \label{eq:gaugesumrule}
\end{equation}
where the last equality is a consequence of completeness of the set of 
states $H^0_i$.

Under our assumption of natural flavor
conservation~\cite{Glashow:1976nt,Paschos:1976ay}, the masses of each
type of fermion are generated by only one doublet.  For fermion
species $f$, the Yukawa Lagrangian can be written in the general form
\begin{equation}
\mathcal{L}_{Yuk} = - y_f \bar f_R \Phi_f^{\dagger} F_L
+ {\rm h.c.},
\end{equation}
where $F_L$ is the appropriate left-handed fermion doublet and
$\Phi_f$ is the Higgs doublet that gives mass to fermion species $f$.  If
$f_R$ is an up-type quark then $\Phi_f^{\dagger}$ should be replaced
by $\tilde \Phi_f^{\dagger}$ above.  These couplings generate the
fermion masses, $m_f = y_f v_f / \sqrt{2} = y_f b_f v_{SM}/\sqrt{2}$.
Note that perturbativity of the Yukawa couplings, 
$y_f \lesssim \sqrt{4 \pi}$, together with the known third-generation 
fermion masses, imposes lower bounds on $b_f$ in each 
fermion sector: $b_u \gtrsim 0.3$, $b_d \gtrsim 0.005$, 
and $b_{\ell} \gtrsim 0.003$.
After the mixing in Eq.~\ref{eq:sasum}, the coupling of the physical
state $h$ to an $f \bar f$ pair is given by
\begin{equation}
g_f^h = \frac{y_f}{\sqrt{2}} \langle h | \phi_f \rangle
= \frac{m_f}{b_f v_{SM}} a_f
= g_f^{SM} \frac{a_f}{b_f}.
\end{equation}

A sum rule can also be constructed for the fermion couplings, as follows:
\begin{equation}
  \sum_{H^0_i} (g_f^{H^0_i})^2 
  = \frac{y_f^2}{2} \sum_{H^0_i} |\langle H^0_i | \phi_f \rangle |^2
  = \frac{y_f^2}{2} = \frac{m_f^2}{b_f^2 v_{SM}^2},
\end{equation}
where we have again used completeness of the set of states $H^0_i$.
Note that, unlike for the gauge coupling sum rule in
Eq.~\ref{eq:gaugesumrule}, the right-hand side is not known a priori
and instead depends on $b_f$.  We will show later that in some models,
$b_f$ can be extracted from the couplings of a single state $h$.

For compactness we define barred couplings normalized to their
corresponding SM values,
\begin{equation}
\bar g_i \equiv \frac{g_i^h}{g_i^{SM}},
\label{eq:gbardef}
\end{equation}
so that $\bar g_i \to 1$ in the SM limit and
\begin{equation}
\bar g_W = \sum_{\rm doublets} a_i b_i, \qquad \qquad
\bar g_f = \frac{a_f}{b_f}.
\end{equation}

\section{Fermion masses from one doublet}
\label{sect:1hdm}

In this section we consider extensions of the Standard Model in which
the masses of all fermions arise from couplings to a single
Higgs doublet.  This class includes models containing additional
doublets that do not couple to the fermions (so-called Type I models),
as well as models containing one or more singlets.

\subsection{Standard Model plus one or more singlets (SM+S)}

The simplest way to extend the Standard Model Higgs sector is to add
one real singlet scalar, $S$, which mixes with the usual SM Higgs boson
to form two CP-even neutral Higgs mass eigenstates.  The constraints of
Eqs.~\ref{eq:sasum}, \ref{eq:sbsum} become
\begin{equation}
a_f^2 + a_s^2 = 1, \qquad \qquad b_f^2 = 1,
\label{eq:sm+s}
\end{equation}
where the subscript $f$ refers to the Higgs doublet, which is
responsible for all fermion masses, and $s$ refers to the singlet.  As
noted below Eq.~\ref{eq:bsum}, $b_f$ can be chosen real and positive,
$b_f = 1$.  Simultaneously, $a_f$ can be chosen real and positive
through an appropriate rephasing of the mass eigenstate $h$; $a_s$ can
then be chosen real and positive through a rephasing of the field $S$.
In particular, we can write
\begin{equation}
a_f = \sqrt{1 - a_s^2} \equiv \sqrt{\xi}.
\end{equation}

The couplings of $h$ to SM particles, normalized to their SM values as
in Eq.~\ref{eq:gbardef}, are then given by
\begin{equation}
\bar g_W = a_f b_f = \sqrt{\xi}, \qquad \qquad
\bar g_f = \frac{a_f}{b_f} = \sqrt{\xi}.
\end{equation}
In particular, the couplings of $h$ to $W$ or $Z$ boson pairs and to
fermion pairs are all scaled down by a common factor $\sqrt{\xi} \leq
1$ relative to their values in the SM.
The production cross sections, partial decay widths, and total width
of $h$ are then all suppressed by a factor of $\xi$ relative to those
of the SM Higgs boson~\footnote{Note that a nonzero branching fraction of $h$ to invisible particles could mimic this effect by suppressing Higgs event rates in all visible channels by a common factor.  This possibility can be tested through a dedicated search for $h \to$ invisible~\cite{Eboli:2000ze,atlas:2003ab,Davoudiasl:2004aj,Barger:2006sk}.}:
\begin{equation}
\frac{\Gamma_i^h}{\Gamma_i^{SM}} = \xi = 1 - a_s^2.
\end{equation}
A measurement of any of these quantities allows a unique determination
of the model parameters $a_f = \sqrt{\xi}$ and $a_s = \sqrt{1 - \xi}$.
Because all the Higgs partial widths scale the same way with $\xi$,
the branching fractions of $h$ are the same as in the SM.

We also note that a decoupling limit can be defined in which $a_s \to 0$
and the couplings of $h$ approach those of the SM Higgs; defining a small
decoupling parameter $\delta \equiv a_s$ we can write
\begin{equation}
  \bar g_f = \bar g_W = \sqrt{1 - \delta^2} \simeq 1 - \frac{1}{2} \delta^2,
  \qquad \qquad
  \frac{\Gamma_i^h}{\Gamma_i^{SM}} = 1 - \delta^2.
  \label{eq:SM+Sdecoup}
\end{equation}

These results can easily be extended to models containing two or more
singlets by making the replacement
\begin{equation}
  a_s^2 \to \sum_{\rm singlets} a_{s_i}^2.
\end{equation}
The relations given above for $a_f$ continue to hold, and we conclude
that models with more than one singlet cannot be distinguished from
the one-singlet model on the basis of the $h$ couplings alone.

\subsection{Type-I Two Higgs Doublet Model (2HDM-I)}

The Type-I 2HDM~\cite{Georgi:1978wr,Haber:1978jt} has been extensively
studied in the literature.  This model contains two scalar SU(2)$_L$
doublets, which we denote by $\Phi_f$ and $\Phi_0$; $\Phi_f$ couples
to fermions and $\Phi_0$ does not.

The constraints of Eqs.~\ref{eq:sasum}, \ref{eq:sbsum} become
\begin{equation}
  a_f^2 + a_0^2 = 1, \qquad \qquad b_f^2 + b_0^2 = 1.
\end{equation}
The couplings of $h$ to SM particles, normalized to their SM values as
in Eq.~\ref{eq:gbardef}, are then given by
\begin{equation}
\bar g_W = a_f b_f + a_0 b_0, \qquad \qquad
\bar g_f = \frac{a_f}{b_f}.
\end{equation}
The vev ratios $b_f$ and $b_0$ can both be chosen real and positive.
Simultaneously, $\bar g_W$ can be chosen real and positive through an
appropriate rephasing of the mass eigenstate $h$.  There is no freedom
left to choose $a_f$ positive, though; thus $\bar g_f$ can have either
sign.

Note in particular that while the couplings of $h$ to fermions of all
three sectors are scaled by the same factor relative to the SM, $\bar
g_u = \bar g_d = \bar g_{\ell}$, the coupling of $h$ to $W$ or $Z$
boson pairs is scaled by a different factor unless $b_0 = 0$.  This
distinguishes the 2HDM-I from the SM plus a singlet discussed above.
In the limit $b_0 \to 0$ we obtain $\bar g_W = \bar g_f = a_f$ and the
couplings of $h$ in the 2HDM-I reduce to those in the SM+S model.

\begin{figure}
\resizebox{0.5\textwidth}{!}{\rotatebox{270}{\includegraphics{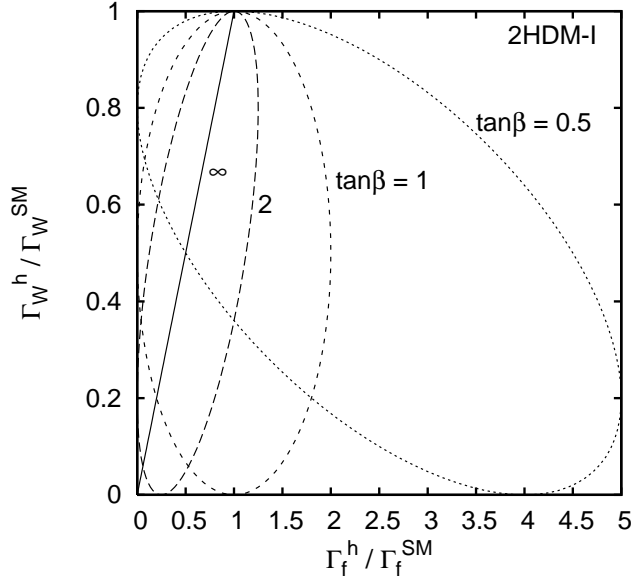}}}
\caption{Surface inhabited by the 2HDM-I in the plane of
$\Gamma_f^h/\Gamma_f^{SM}$ versus $\Gamma_W^h/\Gamma_W^{SM}$.
Here $\tan\beta \equiv v_f/v_0 = b_f/b_0$.  Note
the double covering of the plane for small values of
$\Gamma_f^h/\Gamma_f^{SM}$ and $\Gamma_W^h/\Gamma_W^{SM}$.}
\label{fig:2HDM1}
\end{figure}

The constraint equations and coupling relations can be solved
explicitly for the $b_i$ and $a_i$ factors in terms of the $h$
couplings:
\begin{eqnarray}
b_f &=& \left[ \frac{1 - \bar g_W^2}{1 + \bar g_f^2 - 2 \bar g_W \bar g_f}
  \right]^{1/2}, \qquad \qquad
b_0 = \sqrt{1 - b_f^2}, \nonumber \\
a_f &=& b_f \bar g_f, \qquad \qquad \qquad \qquad \qquad
a_0 = \frac{\bar g_W - b_f^2 \bar g_f}{\sqrt{1 - b_f^2}}.
\end{eqnarray}
Note that a full, unique solution is obtained if the relative signs of
$\bar g_f$ and $\bar g_W$ are known.  If
the relative signs are not known, then there are two possible
solutions, as illustrated in Fig.~\ref{fig:2HDM1}.
Access to the relative signs of the couplings requires
interfering the relevant amplitudes, e.g., by using $h \to
\gamma\gamma$ or $h \to Z \gamma$.

It is useful to make contact with the usual notation for the
2HDM-I~\cite{Gunion:1989we}.  We assume that the state $h$ under
consideration is the lighter of the two neutral CP-even Higgs mass
eigenstates,
\begin{equation}
h^0 = \sqrt{2} \,
(\cos\alpha \, {\rm Re} \, \Phi_f^0 - \sin\alpha \, {\rm Re} \, \Phi_0^0)
= \cos\alpha \, \phi_f - \sin\alpha \, \phi_0,
\end{equation}
with $-\pi/2 < \alpha < \pi/2$, so that $a_f = \cos\alpha$ and $a_0 =
-\sin\alpha$.  The ratio of the doublet vevs can be defined according
to
\begin{equation}
\tan\beta \equiv \frac{v_f}{v_0} = \frac{b_f}{b_0},
\end{equation}
so that $b_f = \sin\beta$ and $b_0 = \cos\beta$.
With these definitions, the couplings of $h$ to SM particles become
\begin{equation}
\bar g_W = \sin(\beta - \alpha), \qquad \qquad
\bar g_f = \frac{\cos\alpha}{\sin\beta}
  = \sin(\beta - \alpha) + \cot\beta \cos(\beta - \alpha).
\end{equation}
We note that for $\tan\beta < 1$ the fermion couplings can be enhanced 
($\bar g_f > 1$).  Perturbativity of the top quark Yukawa coupling
requires $\tan\beta \gtrsim 0.3$; there is no upper bound on $\tan\beta$.
For fixed $\tan\beta$ the maximum possible
value of $\Gamma_f^h/\Gamma_f^{SM}$ is $1 + \cot^2\beta$.

The decoupling limit of this model occurs when the mass eigenstate $h$
coincides with the state $\phi_v$, the vev-carrying doublet in the
Higgs basis.  In that case $a_i \equiv \langle h | \phi_i \rangle
= \langle \phi_v | \phi_i \rangle = b_i$, so that $\bar g_f = a_f/b_f
= 1$ and $\bar g_W = a_f b_f + a_0 b_0 = b_f^2 + b_0^2 = 1$.  Near the
decoupling limit, we can parameterize the deviations of the couplings
from their SM values in terms of a small parameter,
\begin{equation}
  \delta \equiv\cos(\beta - \alpha) = a_f b_0 - a_0 b_f.
\end{equation}
We have, for the couplings of $h$ to gauge bosons,
\begin{equation}
  \bar g_W = \sqrt{1 - \delta^2} \simeq 1 - \frac{1}{2} \delta^2,
  \qquad \qquad
  \frac{\Gamma_W^h}{\Gamma_W^{SM}} = 1 - \delta^2.
  \label{eq:2HDMIdecoup1}
\end{equation}
The couplings of $h$ to fermions depend also on $\tan\beta$:
\begin{equation}
  \bar g_f = \sqrt{1 - \delta^2} + \cot\beta \, \delta
  \simeq 1 + \cot\beta \, \delta - \frac{1}{2} \delta^2,
  \qquad \qquad
  \frac{\Gamma_f^h}{\Gamma_f^{SM}} \simeq 1 + 2 \cot\beta \, \delta
  - \delta^2,
  \label{eq:2HDMIdecoup2}
\end{equation}
where the terms of order $\delta^2$ must be kept if $\cot\beta$ is very
close to zero.  Note that $\delta$ can take either sign.

\subsection{2HDM-I plus one or more singlets (2HDM-I+S)}

We next consider the consequences of adding a real singlet scalar
field, $S$, to the 2HDM-I.  The constraints of Eqs.~\ref{eq:sasum},
\ref{eq:sbsum} become
\begin{equation}
  a_f^2 + a_0^2 + a_s^2 = 1, \qquad \qquad b_f^2 + b_0^2 = 1,
\end{equation}
where $a_s \equiv \langle h | S \rangle$ and the other $a_i$, $b_i$
are defined as in the previous section.  
The couplings of $h$ to SM particles, normalized to their SM
values, are given as for the 2HDM-I by
\begin{equation}
  \bar g_W = a_f b_f + a_0 b_0, \qquad \qquad
  \bar g_f = \frac{a_f}{b_f}.
\end{equation}
As before, $b_f$, $b_0$, and $\bar g_W$ can be chosen real and
positive, while $\bar g_f$ can have either sign.  The coefficient
$a_s$ can then be chosen real and positive by a rephasing of $S$.

With five parameters and only four equations, the parameters of this
model cannot be fully solved for in terms of the $h$ couplings.  In
order to display the ambiguity we define
\begin{equation}
  \xi \equiv 1 - a_s^2 = a_f^2 + a_0^2,
\end{equation}
with $0 < \xi \leq 1$ parameterizing the doublet content of $h$.  We
then obtain,
\begin{eqnarray}
b_f &=& \left[ \frac{\xi - \bar g_W^2}
  {\xi + \bar g_f^2 - 2 \bar g_W \bar g_f} \right]^{1/2}, \qquad \qquad
b_0 = \sqrt{1 - b_f^2}, \nonumber \\
a_f &=& b_f \bar g_f, \qquad \qquad
a_0 = \frac{\bar g_W - b_f^2 \bar g_f}{\sqrt{1 - b_f^2}}, \qquad \qquad
a_s = \sqrt{1 - \xi},
\end{eqnarray}
where $\xi$ remains an undetermined parameter.

This model can be cast into the usual notation of the 2HDM-I as
follows.  We first parameterize the doublet-singlet mixing in
terms of $\xi$,
\begin{equation}
h = \sqrt{\xi} \, h^{\prime} + \sqrt{1 - \xi} \, S,
\end{equation}
where $h^{\prime}$ corresponds to our Higgs in the 2HDM-I in the limit
of zero singlet admixture:
\begin{equation}
h^{\prime} = \cos\alpha \, \phi_f - \sin\alpha \, \phi_0.
\end{equation}
We then have $a_f = \sqrt{\xi} \cos\alpha$, $a_0 = -\sqrt{\xi}
\sin\alpha$, and $a_s = \sqrt{1-\xi}$.  The couplings are given by
\begin{equation}
\bar g_W = \sqrt{\xi} \, \sin(\beta - \alpha), \qquad \qquad
\bar g_f = \sqrt{\xi} \, \frac{\cos\alpha}{\sin\beta}.
\end{equation}
In particular, the couplings of $h$ to SM particles are all scaled
down by a common factor $\sqrt{\xi}$.

Regardless of whether the parameters of the model can be solved for
uniquely in terms of the $h$ couplings, this model would be
distinguishable from the 2HDM-I if it occupied a different
footprint in the space of observables; i.e., if one could obtain
sets of couplings $(\bar g_W, \bar g_f)$ in this model that
could not be obtained in the 2HDM-I.  \emph{This is not the case.}
Any set of couplings $(\bar g_W, \bar g_f)$ that can be
obtained in the 2HDM-I+S can also be obtained in the 2HDM-I, albeit
from different underlying values of the parameters $a_f$, $a_0$,
$b_f$, and $b_0$.  In particular, the models are identical when
$\sqrt{\xi} = 1$; away from this limit the ellipses in
Fig.~\ref{fig:2HDM1} are simply scaled down by $\xi$ on both
axes.

The presence of the singlet thus cannot be established through
measurements of the couplings of $h$ only.  However, if the couplings
of a second CP-even neutral Higgs state ($H$) could be measured,
nonzero singlet mixing would violate the usual 2HDM coupling sum
rule~\cite{Gunion:1990kf}, $(g^h_W)^2 + (g^H_W)^2 = (g^{SM}_W)^2$
(see Eq.~\ref{eq:gaugesumrule}).
This violation would indicate the presence of a third CP-even state
such that $\sum_{i=1}^3 (g^{h_i}_W)^2 = (g^{SM}_W)^2$.

These results can easily be extended to models containing two or more
singlets by making the replacement
\begin{equation}
  a_s^2 \to \sum_{\rm singlets} a_{s_i}^2 = 1 - \xi.
\end{equation}
Again, such a model cannot be distinguished from the 2HDM-I on the
basis of the $h$ couplings alone.

We conclude that adding one or more singlets to the 2HDM-I results in
a model that cannot be distinguished from the 2HDM-I on the basis of
the $h$ couplings alone.

\subsection{2HDM-I plus additional doublet(s) (2HDM-I+D)}
\label{sec:2HDM-I+D}

Let us now consider the consequences of adding one or more additional
Higgs doublets to the 2HDM-I.  These additional doublets can carry
vevs but, under our assumption of natural flavor conservation, they
must not couple to fermions.  We can denote the field content of the
model as $\Phi_f$, ${\Phi_0}_i$, with $i = 1 \ldots n$ ($n \geq 2$)
counting the doublets that do not couple to fermions.

We first define a linear combination $\phi_0^{\prime}$ of the neutral
CP-even states ${\phi_0}_i$ such that
\begin{equation}
  h = a_f \phi_f + a_0^{\prime} \phi_0^{\prime}, \qquad \qquad
  a_f^2 + a_0^{\prime 2} = 1.
\end{equation}
The vev of $\phi_0^{\prime}$ is parameterized by $b_0^{\prime} \equiv
\langle \phi_0^{\prime} | \phi_v \rangle$, chosen to be positive; the
phase of $h$ is chosen to make $\bar g_W = a_f b_f + a_0^{\prime}
b_0^{\prime}$ positive.  Eq.~\ref{eq:sbsum} is modified to read
\begin{equation}
  b_f^2 + b_0^{\prime 2} = \omega^2, \qquad \qquad 0 < \omega \leq 1,
\end{equation}
where $\omega < 1$ indicates that a nonzero vev is carried by the
linear combination(s) of $\phi_{0_i}$ orthogonal to $h$.  Again, this
model has five parameters but only four constraint equations; the
solution for the model parameters becomes
\begin{eqnarray}
b_f &=& \left[ \frac{\omega^2 - \bar g_W^2}
  {1 + \omega^2 \bar g_f^2 - 2 \bar g_W \bar g_f} \right]^{1/2},
\qquad \qquad
b_0^{\prime} = \sqrt{\omega^2 - b_f^2}, \nonumber \\
a_f &=& b_f \bar g_f, \qquad \qquad
a_0^{\prime} = \frac{\bar g_W - b_f^2 \bar g_f}{\sqrt{\omega^2 - b_f^2}},
\end{eqnarray}
where $\omega$ remains an undetermined parameter.

Translating into the usual 2HDM-I notation, we define $\alpha$ as for
the 2HDM-I and $\tan\beta$ as
\begin{equation}
  \tan\beta \equiv \frac{b_f}{b_0^{\prime}},
\end{equation}
where now $\sin\beta = b_f/\omega$ and $\cos\beta =
b_0^{\prime}/\omega$.  In this notation, the couplings of $h$
become
\begin{equation}
  \bar g_W = \omega \sin(\beta - \alpha), \qquad \qquad
  \bar g_f = \frac{1}{\omega} \frac{\cos\alpha}{\sin\beta}.
\end{equation}
The couplings of $h$ to gauge bosons are scaled down by a factor
$\omega \leq 1$ while the couplings of $h$ to fermions are scaled up
by a factor $1/\omega \geq 1$.  Again, this model occupies the same
footprint in Higgs coupling space as the 2HDM-I (consider
Fig.~\ref{fig:2HDM1}), and we conclude that adding one or more
additional doublets to the 2HDM-I results in a model that cannot be
distinguished from the 2HDM-I on the basis of the $h$ couplings alone.
This conclusion remains unchanged if singlets are added to the model
as well.

\section{Fermion masses from two doublets}
\label{sect:2hdm}

We now consider models in which the fermion masses arise from
couplings to two different Higgs doublets.  Imposing natural flavor
conservation allows for three
possible patterns of couplings of two Higgs doublets to the 
fermions~\cite{Barnett:1983mm,Barnett:1984zy,Grossman:1994jb}:
\begin{enumerate}
\item the Type-II 2HDM, or Model II, in which one doublet generates
the masses of the up-type quarks while the other generates the masses
of the down-type quarks and charged leptons (this is the coupling
structure present at tree level in the MSSM);
\item the ``flipped'' 2HDM, in which one doublet generates the masses
of the up-type quarks and charged leptons while the other generates
the masses of the down-type quarks; and
\item the ``lepton-specific'' 2HDM, in which one doublet generates the
masses of up- and down-type quarks while the other generates the
masses of the charged leptons.
\end{enumerate}
We consider here also extensions of these three models obtained by adding
one or more electroweak singlets or doublets that do not couple to
fermions.

While the Higgs sector of the MSSM is a Type-II 2HDM at tree level,
one-loop radiative corrections involving supersymmetric particles can
induce a significant coupling of the bottom quark to the ``wrong''
Higgs doublet, encoded in an extra coupling parameter
$\Delta_b$~\cite{Hall:1993gn,Hempfling:1993kv,Carena:1994bv,Pierce:1996zz}.
This violates our assumption of natural flavor conservation; however,
for completeness, we consider the main features of this model here
separately.

\subsection{Type-II Two Higgs Doublet Model (2HDM-II)}
\label{sec:2HDM-II}

The Type-II 2HDM~\cite{Donoghue:1978cj,Hall:1981bc,Gunion:1989we} is
perhaps the most widely studied extension of the SM Higgs sector.  The
Higgs content and coupling structure are the same as in the MSSM at
tree level.  This model contains two scalar SU(2)$_L$ doublets, which we
denote by $\Phi_u$ and $\Phi_d$.  $\Phi_u$ generates the masses of the
up-type quarks while $\Phi_d$ generates the masses of the down-type
quarks and the charged leptons.

The constraints of Eqs.~\ref{eq:sasum} and \ref{eq:sbsum} become
\begin{equation}
a_u^2 + a_d^2 = 1, \qquad \qquad b_u^2 + b_d^2 = 1.
\label{eq:2HDMIIconstraint}
\end{equation}
The normalized couplings of $h$ to SM particles are then given by
\begin{equation}
  \bar g_W = a_u b_u + a_d b_d, \qquad \qquad
  \bar g_u = \frac{a_u}{b_u}, \qquad \qquad
  \bar g_d = \bar g_{\ell} = \frac{a_d}{b_d},
  \label{eq:coups2HDM2}
\end{equation}
where $\bar g_u$, $\bar g_d$, and $\bar g_{\ell}$ denote the
normalized couplings of $h$ to all three generations of up-type quarks,
down-type quarks, and charged leptons, respectively.

The vev ratios $b_u$ and $b_d$ can both be chosen real and positive.
Simultaneously, we can choose $\bar g_W$ to be real and positive
through an appropriate rephasing of the mass eigenstate $h$.  There is
no freedom left to choose the signs of the fermion couplings $\bar
g_u$ and $\bar g_d$; depending on the underlying values of the
parameters they can take the signs $++$, $+-$, or $-+$; $\bar g_u$ and
$\bar g_d$ cannot both be negative.

The 2HDM-II has four parameters related by five constraints, resulting
in a \emph{pattern relation}~\cite{Ginzburg:2001ss,Ginzburg:2001wj}
among the three couplings of $h$:\footnote{One can define an
additional pattern relation involving $\bar g_W$, $\bar g_u$ and $\bar
g_{\ell}$ according to $P_{u\ell} \equiv \bar g_W (\bar g_u + \bar
g_{\ell}) - \bar g_u \bar g_{\ell} = 1 = P_{ud}$.}
\begin{equation}
  P_{ud} \equiv \bar g_W (\bar g_u + \bar g_d) - \bar g_u \bar g_d = 1.
  \label{eq:pattreln}
\end{equation}
This pattern relation provides a test of the 2HDM-II coupling
structure:
It defines a two-dimensional surface accessible by the
model in the three-dimensional space of couplings $\bar g_W$, $\bar
g_u$, and $\bar g_d$.

The constraint equations and coupling relations can be solved explicitly
for the $b_i$ factors in terms of the $h$ couplings:
\begin{eqnarray}
b_u &=& \left[ \frac{\bar g_W - \bar g_d}{\bar g_u - \bar g_d} \right]^{1/2}
= \left[ \frac{1 - \bar g_d^2}{\bar g_u^2 - \bar g_d^2} \right]^{1/2},
\nonumber \\
b_d &=& \left[ \frac{\bar g_W - \bar g_u}{\bar g_d - \bar g_u} \right]^{1/2}
= \left[ \frac{1 - \bar g_u^2}{\bar g_d^2 - \bar g_u^2} \right]^{1/2},
\label{eq:2HDMIIbubd}
\end{eqnarray}
where in the second relation the dependence on $\bar g_W$ has been removed
using the pattern relation.  We also obtain the $a_i$ factors,
\begin{equation}
  a_u = b_u \bar g_u, \qquad \qquad a_d = b_d \bar g_d.
\end{equation}
If the relative signs of $\bar g_u$, $\bar g_d$, and $\bar
g_W$ are known, then the solution for the model parameters is unique.
Note also that, unlike in the 2HDM-I, a unique solution for $b_u$
and $b_d$ can be obtained even if the signs of the couplings are
not known, by using the second set of equalities in Eq.~\ref{eq:2HDMIIbubd}.
In the absence of information on the signs of the couplings, the
magnitudes of $a_u$ and $a_d$ can also be determined uniquely but
their relative signs cannot.

Let us now make contact with the usual notation for the
2HDM-II~\cite{Gunion:1989we}.  We assume that the state $h$ under
consideration is the lighter of the two neutral CP-even Higgs mass
eigenstates,
\begin{equation}
  h^0 = \sqrt{2} \,
  (\cos\alpha \, {\rm Re} \, \Phi_u^0 - \sin\alpha \, {\rm Re} \, \Phi_d^0),
  \label{eq:2HDM2alphadefinition}
\end{equation}
so that $a_u = \cos\alpha$ and $a_d =
-\sin\alpha$.  The ratio of the doublet vevs can be defined according
to $\tan\beta \equiv v_u/v_d = b_u/b_d$, so that $b_u = \sin\beta$ and
$b_d = \cos\beta$.\footnote{The constraint $\bar g_W \geq 0$ corresponds to
$\beta - \pi \leq \alpha \leq \beta$.  Perturbativity of the top
quark Yukawa coupling requires $\tan\beta \gtrsim 0.3$; perturbativity
of the bottom quark Yukawa coupling requires $\tan\beta \lesssim 200$.}
With these definitions, the couplings of $h$ to SM particles become
\begin{eqnarray}
\bar g_W &=& \sin(\beta - \alpha), \nonumber \\
\bar g_u &=& \frac{\cos\alpha}{\sin\beta}
  = \sin(\beta - \alpha) + \cot\beta \cos(\beta - \alpha), \nonumber \\
\bar g_d &=& \bar g_{\ell} = -\frac{\sin\alpha}{\cos\beta}
  = \sin(\beta - \alpha) - \tan\beta \cos(\beta - \alpha).
\end{eqnarray}
We note that $\tan\beta$ can be obtained from coupling
measurements using
\begin{equation}
\tan\beta = \left[ \frac{\bar g_d^2 - 1}{1 - \bar g_u^2} \right]^{1/2}
= \left[ \frac{\Gamma_d^h/\Gamma_d^{SM} - 1}{1 - \Gamma_u^h/\Gamma_u^{SM}}
    \right]^{1/2}.
\end{equation}

\begin{figure}
\resizebox{0.48\textwidth}{!}{\rotatebox{270}{\includegraphics{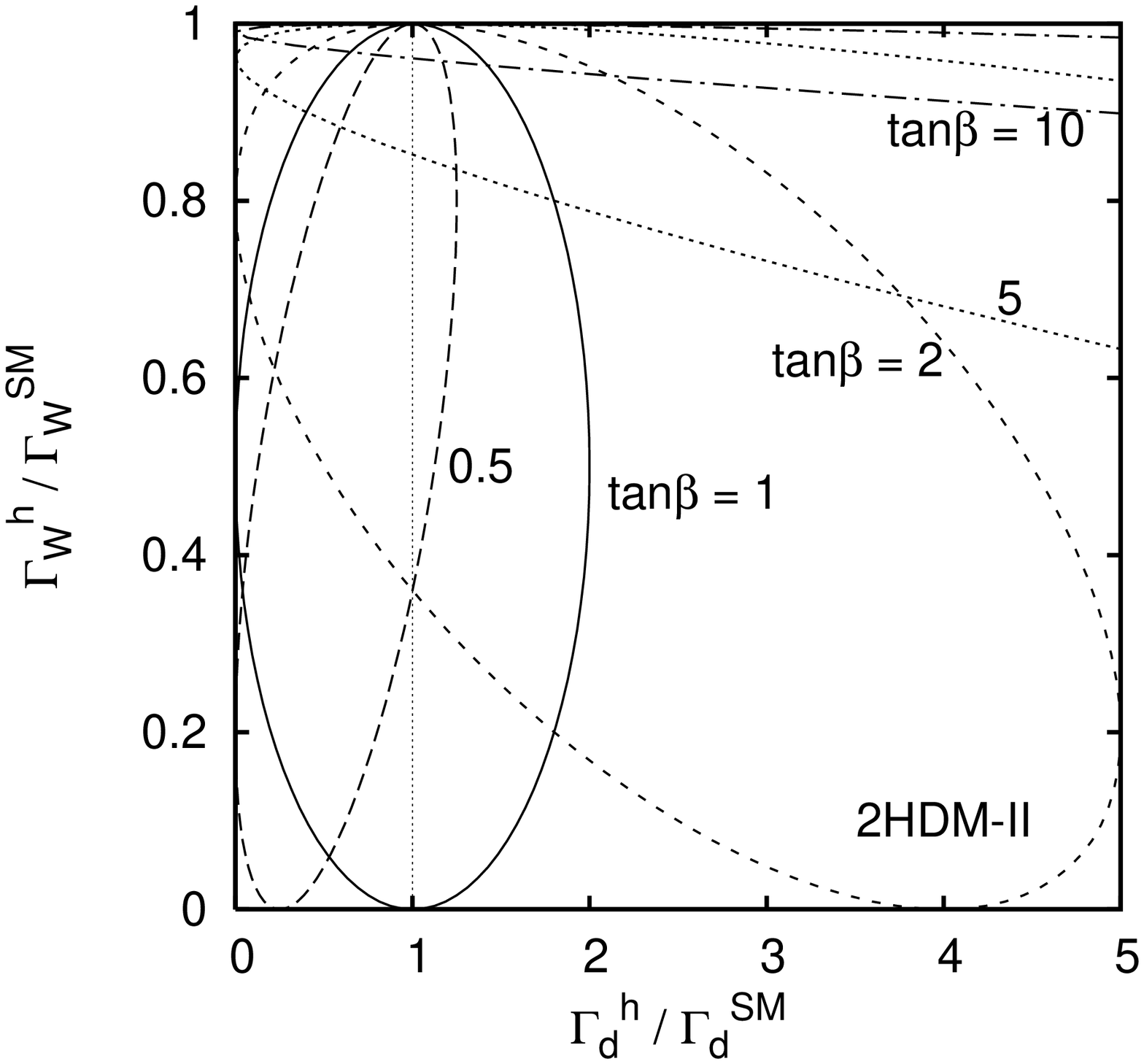}}}
\resizebox{0.48\textwidth}{!}{\rotatebox{270}{\includegraphics{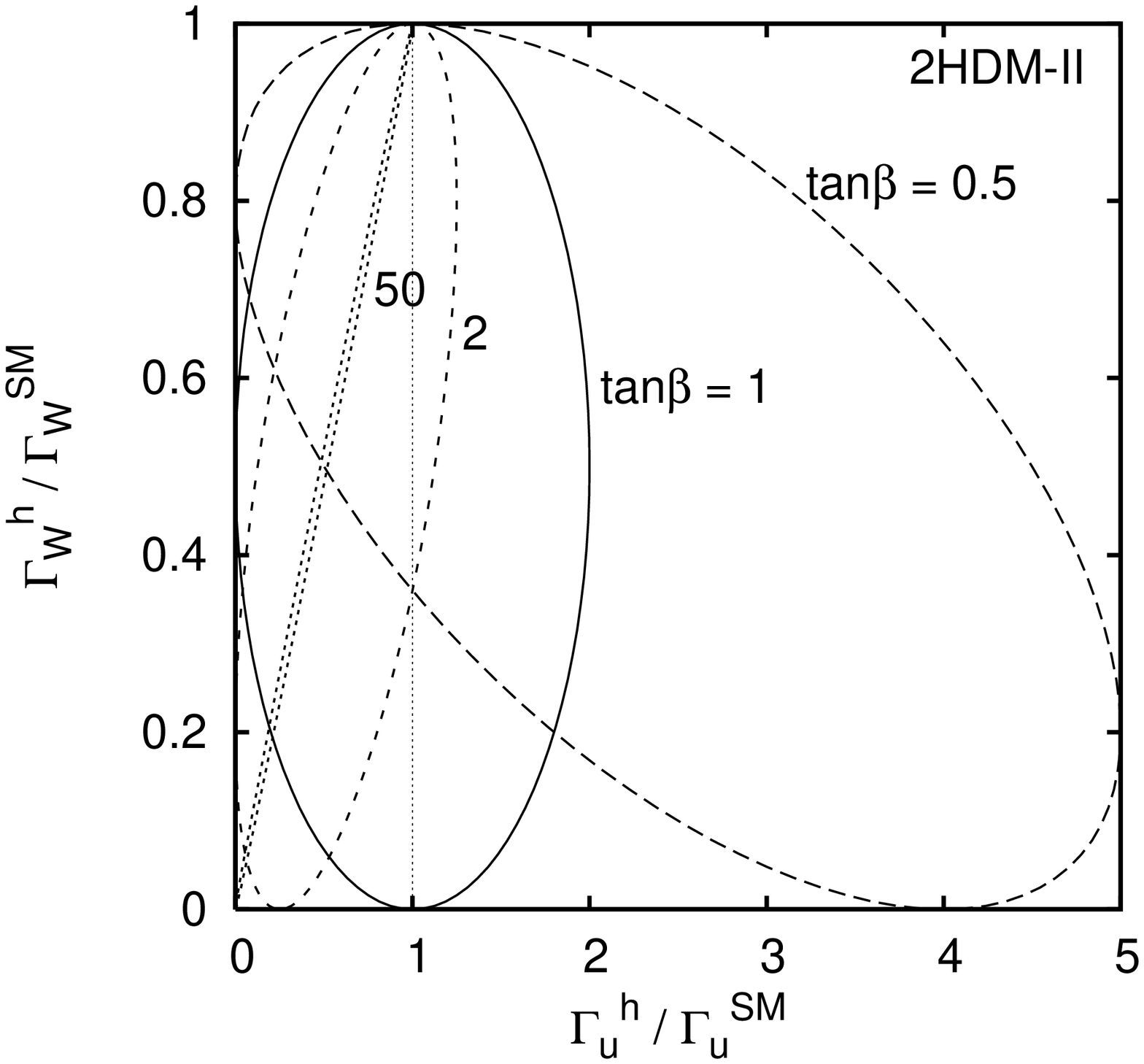}}}
\caption{Surface inhabited by the 2HDM-II in the plane of (left)
$\Gamma_d^h/\Gamma_d^{SM}$ and (right) $\Gamma_u^h/\Gamma_u^{SM}$
versus $\Gamma_W^h/\Gamma_W^{SM}$, for various values of $\tan\beta$.}
\label{fig:2HDM2}
\end{figure}

\begin{figure}
\resizebox{0.48\textwidth}{!}{\rotatebox{270}{\includegraphics{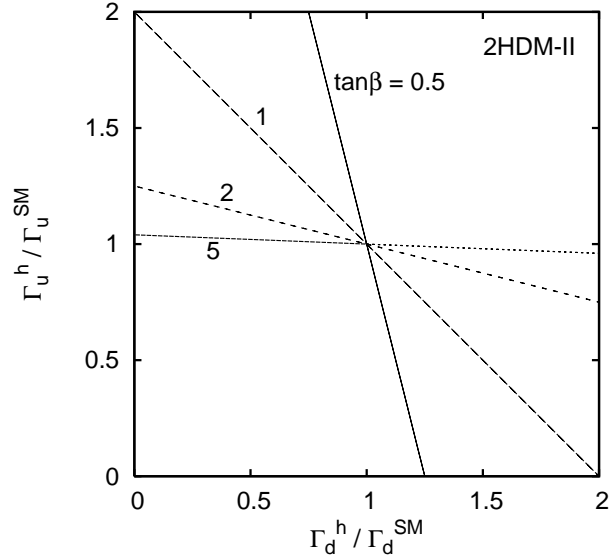}}}
\caption{Surface inhabited by the 2HDM-II in the plane of
$\Gamma_d^h/\Gamma_d^{SM}$ versus $\Gamma_u^h/\Gamma_u^{SM}$.}
\label{fig:2HDM2-d}
\end{figure}

The relations among the Higgs partial widths into up-type quark,
down-type quark (or charged lepton), and $W$ boson pair final states
are shown in Figs.~\ref{fig:2HDM2} and \ref{fig:2HDM2-d} for various
values of $\tan\beta$.  The key difference between the 2HDM-II and the
2HDM-I is the different behavior of $\bar g_u$ compared to $\bar g_d$
(and $\bar g_{\ell}$), as illustrated in Fig.~\ref{fig:2HDM2-d}.  In
particular, the 2HDM-I would fall on a line of slope $+1$ through the
SM point $(1,1)$ on this plot.

The decoupling limit of this model occurs when the mass eigenstate $h$
coincides with the state $\phi_v$, the vev-carrying doublet in the
Higgs basis.  Near the decoupling limit we parameterize the
deviations of the couplings from their SM values in terms of a small
parameter
\begin{equation}
  \delta \equiv \cos(\beta - \alpha) = a_u b_d - a_d b_u.
\end{equation}
The couplings and corresponding partial widths, normalized to their SM
values, become
\begin{eqnarray}
  &\bar g_W = \sqrt{1 - \delta^2} \simeq 1 - \frac{1}{2} \delta^2,
  \qquad \qquad
  &\frac{\Gamma_W^h}{\Gamma_W^{SM}} = 1 - \delta^2, \nonumber \\
  &\bar g_u = \sqrt{1 - \delta^2} + \cot\beta \, \delta
  \simeq 1 + \cot\beta \, \delta,
  \qquad \qquad
  &\frac{\Gamma_u^h}{\Gamma_u^{SM}} \simeq 1 + 2 \cot\beta \, \delta,
  \nonumber \\
  &\bar g_d = \bar g_{\ell} = \sqrt{1 - \delta^2} - \tan\beta \, \delta
  \simeq 1 - \tan\beta \, \delta,
  \qquad \qquad
  &\frac{\Gamma_d^h}{\Gamma_d^{SM}} \simeq 1 - 2 \tan\beta \, \delta.
  \label{eq:2HDMIIdecoup}
\end{eqnarray}
Note that $\delta$ can take either sign.

\subsection{2HDM-II plus one or more singlets (2HDM-II+S)}

We now consider the consequences of adding a real singlet scalar field,
$S$, to the 2HDM-II.  The constraints of Eqs.~\ref{eq:sasum},
\ref{eq:sbsum} become
\begin{equation}
  a_u^2 + a_d^2 + a_s^2 = 1, \qquad \qquad b_u^2 + b_d^2 = 1,
\end{equation}
where $a_s \equiv \langle h | S \rangle$ and the other $a_i$, $b_i$
are defined as in Sec.~\ref{sec:2HDM-II}.  The couplings of $h$ to SM
particles, normalized to their SM values, are given in terms of
$a_{u,d}$ and $b_{u,d}$ as for the 2HDM-II by Eq.~\ref{eq:coups2HDM2}.
As before, $b_u$, $b_d$, and $\bar g_W$ can be chosen real and
positive.  The coefficient $a_s$ can be chosen real and positive by a
rephasing of $S$.

Because of the presence of the additional parameter $a_s$, the pattern
relation of the 2HDM-II no longer holds.  Instead we obtain
\begin{equation}
  P_{ud} \equiv \bar g_W (\bar g_u + \bar g_d) - \bar g_u \bar g_d
  = \xi \leq 1,
  \label{eq:pattrel+S}
\end{equation}
where $\xi \equiv 1 - a_s^2 = a_u^2 + a_d^2$ parameterizes the doublet content of $h$.  In particular, $\xi$ can
be determined by applying the pattern relation to measurements of the
couplings $\bar g_W$, $\bar g_u$ and $\bar g_d$.  The solutions for
the rest of the model parameters then become
\begin{eqnarray}
  b_u &=& \left[ \frac{\bar g_W - \bar g_d}{\bar g_u - \bar g_d} \right]^{1/2}
  = \left[ \frac{\xi - \bar g_d^2}{\bar g_u^2 - \bar g_d^2} \right]^{1/2},
  \nonumber \\
  b_d &=& \left[ \frac{\bar g_W - \bar g_u}{\bar g_d - \bar g_u} \right]^{1/2}
  = \left[ \frac{\xi - \bar g_u^2}{\bar g_d^2 - \bar g_u^2} \right]^{1/2},
  \nonumber \\
  a_u &=& b_u \bar g_u, \qquad \qquad a_d = b_d \bar g_d,
  \qquad \qquad a_s = \sqrt{1 - \xi},
\end{eqnarray}
where in the expressions for $b_u$, $b_d$ we have shown how the
dependence on $\bar g_W$ can be traded for dependence on $\xi$ using
Eq.~\ref{eq:pattrel+S}.

Clearly, if the relative signs of $\bar g_W$, $\bar g_u$ and $\bar
g_d$ are known, then this model can be distinguished from the 2HDM-II
using the pattern relation ($P_{ud}$ is equal to one in the 2HDM-II
and less than one in the 2HDM-II+S) and the solution for the model
parameters is unique.  If, however, the relative signs of $\bar g_W$,
$\bar g_u$ and $\bar g_d$ are not known, then $\xi$ cannot be obtained
uniquely and there will be discrete ambiguities in the solutions for
all the parameters.  In this situation the pattern relation can still
be used to test for the presence of the singlet in the model; if no
combination of signs of the Higgs couplings gives $P_{ud} = 1$, then
the model cannot be the minimal 2HDM-II.

This model can be cast into the usual notation for the 2HDM-II as
follows.  We first parameterize the doublet-singlet mixing in terms
of $\xi$,
\begin{equation}
  h = \sqrt{\xi} \, h^{\prime} + \sqrt{1 - \xi} \, S,
\end{equation}
where $h^{\prime}$ corresponds to the Higgs state considered in the
2HDM-II in the limit of zero singlet admixture:
\begin{equation}
  h^{\prime} = \cos\alpha \, \phi_u - \sin\alpha \, \phi_d.
\end{equation}
We then have $a_u = \sqrt{\xi} \, \cos\alpha$ and $a_d = -\sqrt{\xi} \,
\sin\alpha$.  The couplings are given by
\begin{equation}
  \bar g_W = \sqrt{\xi} \, \sin(\beta - \alpha), \qquad \qquad
  \bar g_u = \sqrt{\xi} \, \frac{\cos\alpha}{\sin\beta}, \qquad \qquad
  \bar g_d = \bar g_{\ell} = - \sqrt{\xi} \, \frac{\sin\alpha}{\cos\beta}.
\end{equation}
In particular, the couplings of $h$ to SM particles are all scaled
down by a common factor $\sqrt{\xi} \leq 1$.  This means that the
2HDM-II+S lives on a \emph{volume} in the three-dimensional parameter
space of $\bar g_W$, $\bar g_u$, and $\bar g_d$, consisting of the
surface inhabited by the 2HDM-II (corresponding to $\xi = 1$) together
with all lines that connect points on that surface to the origin
(corresponding to $0 \leq \xi < 1$).  Clearly, the 2HDM-II+S occupies
a different footprint in coupling space than the 2HDM-II, and it can
thus be distinguished from the 2HDM-II.  This is different from the
case of the 2HDM-I+S; the reason is that the Type-II fermion coupling
structure yields a third observable coupling related nontrivially to
the other two.

The decoupling limit comprises $\delta \equiv \cos(\beta - \alpha) =
a_u b_d - a_d b_u \to 0$ and $\epsilon \equiv \sqrt{1 - \xi} \to 0$.
(Note that $\delta$ can have either sign while $\epsilon$ is chosen
positive.)  The couplings and corresponding partial widths, normalized
to their SM values, become
\begin{eqnarray}
  &\bar g_W = \sqrt{1 - \delta^2} \sqrt{1 - \epsilon^2}
  \simeq 1 - \frac{1}{2} \delta^2 - \frac{1}{2} \epsilon^2,
  &\frac{\Gamma_W^h}{\Gamma_W^{SM}} \simeq 1 - \delta^2 - \epsilon^2
  \\
  &\bar g_u = \left[ \sqrt{1 - \delta^2} + \cot\beta \, \delta \right]
  \sqrt{1 - \epsilon^2}
  \simeq 1 + \cot\beta \, \delta - \frac{1}{2} \epsilon^2,
  &\frac{\Gamma_u^h}{\Gamma_u^{SM}} \simeq 1 + 2 \cot\beta \, \delta
  - \epsilon^2, \nonumber \\
  \bar g_d = &\bar g_{\ell} = \left[ \sqrt{1 - \delta^2}
  - \tan\beta \, \delta \right] \sqrt{1 - \epsilon^2}
  \simeq 1 - \tan\beta \, \delta - \frac{1}{2} \epsilon^2, \qquad
  &\frac{\Gamma_d^h}{\Gamma_d^{SM}} \simeq 1 - 2 \tan\beta \, \delta
  - \epsilon^2. \nonumber
\end{eqnarray}
Note that in the limit $\delta \to 0$ with $\epsilon$ finite, the
deviations of the $h$ couplings from their SM values become identical
to those in the SM+S.

These results can easily be extended to models containing two or more
singlets by making the replacement
\begin{equation}
  a_s^2 \to \sum_{\rm singlets} {a_s}_i^2 = 1 - \xi.
\end{equation}
Such a model cannot be distinguished from the 2HDM-II+S (with only one
singlet) on the basis of the $h$ couplings alone.

\subsection{2HDM-II plus additional doublet(s) (2HDM-II+D)}
\label{sec:2HDM-II+D}

We now consider the consequences of adding an additional Higgs doublet
$\Phi_0$ to the 2HDM-II.  The additional doublet can carry a vev, but
under our assumption of natural flavor conservation it must not couple
to fermions.  The constraint equations become,
\begin{equation}
  a_u^2 + a_d^2 + a_0^2 = 1, \qquad \qquad
  b_u^2 + b_d^2 + b_0^2 = 1,
  \label{eq:2HDMIID}
\end{equation}
where $a_0 \equiv \langle h | \phi_0 \rangle$ and $b_0 \equiv
v_0/v_{SM}$.  The normalized couplings of $h$ to SM particles are
given by
\begin{equation}
  \bar g_W = a_u b_u + a_d b_d + a_0 b_0, \qquad \qquad
  \bar g_u = \frac{a_u}{b_u}, \qquad \qquad
  \bar g_d = \bar g_{\ell} = \frac{a_d}{b_d}.
\end{equation}
All three $b_i$ parameters can be chosen real and positive; $\bar g_W$
can also be chosen positive through an appropriate rephasing of $h$.
Any combination of signs is then possible for $\bar g_u$ and $\bar
g_d$; in particular, both can be negative (for $a_0 b_0 > |a_u b_u +
a_d b_d|$) in contrast to the 2HDM-II or 2HDM-II+S.

Because of the presence of the two additional parameters $a_0$ and
$b_0$, the model is underconstrained and the parameters $a_i, b_i$
cannot be extracted in terms of the $h$ couplings.  However, the
model \emph{is} distinguishable from the 2HDM-II because the pattern relation
of Eq.~\ref{eq:pattreln} no longer holds.  In some parts of parameter
space, this model can also be distinguished from the 2HDM-II+S.

In order to illustrate these features, we cast the model into the
usual notation for the 2HDM-II.  We first parameterize the mixing of
the third doublet in terms of an angle $\theta$,
\begin{equation}
  h = \cos\theta \, h^{\prime} + \sin\theta \, \phi_0,
\end{equation}
where $h^{\prime} \equiv \cos\alpha \, \phi_u - \sin\alpha \, \phi_d$
corresponds to the Higgs in the 2HDM-II in the limit of zero mixing
with the extra doublet.  We then have $a_u = \cos\theta \cos\alpha$,
$a_d = -\cos\theta \sin\alpha$, and $a_0 = \sin\theta$.  We also
define $\tan\beta \equiv v_u/v_d = b_u/b_d$ and $\cos\Omega \equiv
\sqrt{b_u^2 + b_d^2}$, $\sin\Omega = b_0$, where the angle $0 \leq
\Omega < \pi/2$ parameterizes the amount of vev carried by $\Phi_0$.
The couplings of $h$ are then given by
\begin{eqnarray}
  \bar g_W &=& \cos\Omega \, \cos\theta \, \sin(\beta - \alpha)
  + \sin\Omega \, \sin\theta, \nonumber \\
  \bar g_u &=& \frac{\cos\theta}{\cos\Omega} \,
  \frac{\cos\alpha}{\sin\beta}, \qquad \qquad
  \bar g_d = \bar g_{\ell} = -\frac{\cos\theta}{\cos\Omega} \,
  \frac{\sin\alpha}{\cos\beta}.
\end{eqnarray}
We note the features of two limiting cases:
\begin{enumerate}
\item When $b_0 = 0$ (i.e., $\cos\Omega = 1$), the $h$ couplings
reduce to those of the 2HDM-II+S, with $\sqrt{\xi}$ replaced by
$\cos\theta$.  This happens because in this limit, $\phi_0$ does not
couple to fermion pairs or gauge boson pairs and the physics is simply
that of the 2HDM-II with mixing of a ``sterile'' state into $h$.  The
pattern relation in this special case becomes
\begin{equation}
  P_{ud} \equiv \bar g_W (\bar g_u + \bar g_d) - \bar g_u \bar g_d
  = \cos^2\theta \leq 1.
\end{equation}
(Note that the 2HDM-II result $P_{ud} = 1$ is recovered in the limit
$\cos^2\theta \to 1$.)
\item When $a_0 = 0$ (i.e., $\cos\theta = 1$, so $h = h^{\prime}$)
there is no mixing of the new doublet into $h$, but the vev of
$\phi_0$ is nonzero so that the total vev carried by the two doublets
that couple to fermions is reduced.  The fermion Yukawa couplings must
thus be enhanced in order to yield the required fermion masses, while
the coupling of $h$ to $W$ or $Z$ pairs is suppressed.  In this case
the couplings of $h$ become
\begin{equation}
  \bar g_W = \cos\Omega \, \sin(\beta - \alpha), \qquad \qquad \bar
  g_u = \frac{1}{\cos\Omega} \, \frac{\cos\alpha}{\sin\beta}, \qquad
  \qquad \bar g_d = -\frac{1}{\cos \Omega} \,
  \frac{\sin\alpha}{\cos\beta},
\end{equation}
and the pattern relation in this special case becomes
\begin{equation}
  P_{ud} \equiv \bar g_W (\bar g_u + \bar g_d) - \bar g_u \bar g_d =
  1 + \tan^2\Omega
  \frac{\sin\alpha \cos\alpha}{\sin\beta \cos\beta}.
  \label{eq:pattrel2HDMIID}
\end{equation}
In particular, $P_{ud} > 1$ whenever $\sin\alpha \cos\alpha > 0$,
i.e., whenever $\bar g_d$ and $\bar g_u$ have opposite signs.
Furthermore, when $\sin\alpha \cos\alpha < 0$ (i.e., when $\bar g_d$
and $\bar g_u$ have the same sign), small values of $\cos\Omega$,
$\sin\beta$, and/or $\cos\beta$ can yield $P_{ud} < 0$.  Either of
these situations allows the 2HDM-II+D to be distinguished from both the
2HDM-II and the 2HDM-II+S.  (Note that the 2HDM-II result $P_{ud} = 1$
is recovered in the limit $\cos\Omega \to 1$.)
\end{enumerate}

In the general case of both $\cos\theta < 1$ (nonzero mixing of
$\phi_0$ into $h$) and $\cos\Omega < 1$ (nonzero vev of $\phi_0$), the
footprint of the 2HDM-II+D covers a three-dimensional volume in the
space of couplings $(\bar g_W, \bar g_u, \bar g_d)$.  Part of this
volume overlies the footprint of the 2HDM-II+S (when mixing dominates,
or when $\bar g_u$ and $\bar g_d$ have the same sign), but part is
unique to the 2HDM-II+D (when vev sharing dominates, or when $\bar
g_u$ and $\bar g_d$ are both negative as discussed before).  Thus the
model is distinguishable in general from the 2HDM-II, and is
distinguishable from the 2HDM-II+S in some regions of parameter space.

We now describe the approach to decoupling in this model.  The
decoupling limit corresponds to $\langle h | \phi_v \rangle \to 1$.
Deviations from this limit can be parameterized by writing 
\begin{equation}
  h = c_{\parallel} \phi_v + c_{\perp} \phi_{\perp}, \qquad \qquad 
  \langle \phi_{\perp} | \phi_v \rangle = 0,
\end{equation}
where $c_{\parallel}^2 + c_{\perp}^2 = 1$ and $\phi_v$ is given in our
parameterization by
\begin{equation}
  \phi_v = \cos\Omega \, (\sin\beta \, \phi_u + \cos\beta \, \phi_d) 
  + \sin\Omega \, \phi_0.
\end{equation}
The component of $h$ orthogonal to $\phi_v$ can be constructed as follows.
We first define two states orthogonal to $\phi_v$ and to each other:
\begin{equation}
  \phi_{\perp 1} = \cos\beta \, \phi_u - \sin\beta \, \phi_d,
\end{equation}
which lies in the $\phi_u$--$\phi_d$ plane, and
\begin{equation}
  \phi_{\perp 2} = -\sin\Omega \, (\sin\beta \, \phi_u + \cos\beta \, \phi_d) 
  + \cos\Omega \, \phi_0.
\end{equation}
Then $\phi_{\perp}$ can be parameterized in terms of a new mixing
angle $\gamma$,
\begin{eqnarray}
  \phi_{\perp} &=& \sin\gamma \, \phi_{\perp 1} + \cos\gamma \, \phi_{\perp 2} 
  \nonumber \\
  &=& (\sin\gamma \cos\beta - \cos\gamma \sin\Omega \sin\beta) \, \phi_u
  + (-\sin\gamma \sin\beta - \cos\gamma \sin\Omega \cos\beta) \, \phi_d 
  \nonumber \\
  &&	+ (\cos\gamma \cos\Omega) \, \phi_0.
  \label{eq:gammadef}
\end{eqnarray}

Defining the decoupling parameter $\delta \equiv c_{\perp}$, we obtain 
the couplings of $h$:
\begin{eqnarray}
  \bar g_W &=& \langle h | \phi_v \rangle = \sqrt{1 - \delta^2} \nonumber \\
  \bar g_u &=& \frac{\langle h | \phi_u \rangle}{b_u} = 
  \sqrt{1 - \delta^2} 
  + \delta \left[ \sin\gamma \frac{\cot\beta}{\cos\Omega} 
    - \cos\gamma \tan\Omega \right] \nonumber \\
  \bar g_d = \bar g_\ell &=& \frac{\langle h | \phi_d \rangle}{b_d} 
  = \sqrt{1 - \delta^2} 
  + \delta \left[ -\sin\gamma \frac{\tan\beta}{\cos\Omega} 
    - \cos\gamma \tan\Omega \right]. 
  \label{eq:decoup2HDM2D}
\end{eqnarray}
Letting $\delta$ take either sign, we can fix $0 \leq \gamma < \pi$.
Note that for $\sin\gamma = \cos\Omega = 1$, these formulas reduce to
those for the 2HDM-II given in Eq.~\ref{eq:2HDMIIdecoup}.
The decoupling limit corresponds to $\delta \to 0$.

This analysis can be extended to the 2HDM-II plus two or more doublets
in a straightforward way.  We denote the doublets that do not couple
to fermions as ${\Phi_0}_i$, with $i = 1 \ldots n$ ($n \geq 2$).  As
in Sec.~\ref{sec:2HDM-I+D}, we first define a linear combination
$\phi_0^{\prime}$ of the neutral CP-even states $\phi_{0_i}$ such that
\begin{equation}
  h = a_u \phi_u + a_d \phi_d + a_0^{\prime} \phi_0^{\prime}, \qquad \qquad
  a_u^2 + a_d^2 + a_0^{\prime 2} = 1.
  \label{eq:2HDMIIDext}
\end{equation}
The vev of $\phi_0^{\prime}$ is parameterized by $b_0^{\prime} \equiv
\langle \phi_0^{\prime} | \phi_v \rangle$, and Eq.~\ref{eq:sbsum}
becomes $b_u^2 + b_d^2 + b_0^{\prime 2} \leq 1$, where the inequality
accounts for the vev carried by the linear combination(s) of
$\phi_{0_i}$ orthogonal to $h$.  While the underlying physics of this
model differs from that of the 2HDM-II plus one extra doublet, the
footprint of the model in coupling space is the same.  This can be
seen straightforwardly by noting that the pattern relation $P_{ud}$
can take any value in the 2HDM-II+D, leaving no room for a larger
footprint when additional extra doublets are added.  Thus, while it is
possible to tell from the couplings of $h$ alone that (at least) one
additional doublet has been added to the 2HDM-II, it is not possible
to tell how many.

The addition of singlet(s) to the 2HDM-II+D can be parameterized in a
similar way.  We first note that as far as the couplings of $h$ are
concerned, adding an additional Higgs doublet with zero vev is
indistinguishable from adding a singlet.  We again obtain
Eq.~\ref{eq:2HDMIIDext} in which $\phi_0^{\prime}$ now denotes the
appropriate linear combination of $\phi_0$ and the singlets.  The
constraint equation for the doublet vevs remains as given in
Eq.~\ref{eq:2HDMIID}.  We see that the model occupies the same
footprint in $h$ coupling space as the 2HDM-II+D and thus it is not
possible on the basis of $h$ couplings alone to tell whether the
2HDM-II+D also contains additional singlets.

\subsection{Flipped 2HDM, lepton-specific 2HDM, and their extensions}

The flipped and lepton-specific two Higgs doublet models comprise the
two possible alternate assignments of fermion couplings of the
two-doublet models considered here.  These models were introduced in
Refs.~\cite{Barnett:1983mm,Barnett:1984zy,Grossman:1994jb}.  Some early studies of their
phenomenology have been made in Refs.~\cite{AkeroydLEP, Akeroyd:1998ui}.  Much can be
extrapolated in a straightforward way from existing results for the
usual 2HDM-I and 2HDM-II.

In the flipped 2HDM, one doublet $\Phi_u$ generates the masses of the
up-type quarks \emph{and the charged leptons} while the other doublet
$\Phi_d$ generates the masses of the down-type quarks.  The constraint
equations remain identical to those of the 2HDM-II as given in
Eq.~\ref{eq:2HDMIIconstraint}, while the normalized couplings of $h$
to SM particles are given by
\begin{equation}
\bar g_W = a_u b_u + a_d b_d, \qquad \qquad
\bar g_u = \bar g_{\ell} = \frac{a_u}{b_u}, \qquad \qquad
\bar g_d = \frac{a_d}{b_d}.
\end{equation}
The distinguishing feature of this model is the behavior of $\bar
g_{\ell}$.  The quark coupling results carry over unchanged from the
2HDM-II model and its extensions by extra doublets and/or singlets.

In the lepton-specific 2HDM\footnote{LHC phenomenology for $h$ in the
lepton-specific 2HDM was also discussed in Ref.~\cite{BrooksThomasPheno08}.
Ref.~\cite{Aoki:2008av} also makes use of this fermion coupling
structure.}, one doublet $\Phi_q$ generates the masses of all flavors
of quarks while the other doublet $\Phi_{\ell}$ generates the masses
of the charged leptons.  The constraint equations become
\begin{equation}
a_q^2 + a_{\ell}^2 = 1,
\qquad \qquad
b_q^2 + b_{\ell}^2 = 1.
\end{equation}
The normalized couplings of $h$ to SM particles are given by
\begin{equation}
\bar g_W = a_q b_q + a_{\ell} b_{\ell}, \qquad \qquad
\bar g_u = \bar g_d = \frac{a_q}{b_q}, \qquad \qquad
\bar g_{\ell} = \frac{a_{\ell}}{b_{\ell}}.
\end{equation}
Note that in the quark sector, this model is identical to the 2HDM-I.
Its distinguishing feature, however, is again the behavior of $\bar
g_{\ell}$; the pattern relation and all other results for the 2HDM-II
and its extensions by extra doublets and/or singlets carry over to
this model with the replacements
\begin{equation}
\bar g_u \rightarrow \bar g_q, \qquad \qquad
\bar g_d \rightarrow \bar g_{\ell}.
\end{equation}

\subsection{MSSM (2HDM-II with $\Delta_b$)}

At tree level, the Higgs sector of the MSSM is a Type-II 2HDM.  The
natural flavor conservation structure of the Yukawa couplings is
enforced by the analyticity of the superpotential.  Beyond tree level,
however, radiative corrections involving loops of supersymmetric
particles can induce additional couplings of right-handed fermions to
the ``wrong'' Higgs
doublet~\cite{Hall:1993gn,Hempfling:1993kv,Carena:1994bv,Pierce:1996zz}.
Thus, beyond tree level the Higgs sector of the MSSM is technically a
Type III 2HDM\footnote{The phenomenology of the general Type III 2HDM
has been reviewed in Ref.~\cite{Atwood:1996vj}.  In this model the
basis chosen for the two Higgs doublets is somewhat arbitrary;
basis-independent methods have been developed in
Refs.~\cite{Davidson:2005cw,Haber:2006ue}}.  This violation of natural
flavor conservation is a consequence of supersymmetry breaking;
because of this, the loop-induced wrong-Higgs couplings do not
decouple as all SUSY mass parameters are simultaneously taken
large~\cite{Carena:1998gk,Haber:2000kq}.

The most important loop-induced wrong Higgs couplings of this type
arise in the bottom-quark sector from loops involving bottom squarks
and gluinos (involving the large QCD gauge coupling) and from loops
involving top squarks and charginos (involving the large top Yukawa
coupling).  In particular, the effect of the wrong-Higgs coupling on
$\bar g_b$ is enhanced by $\tan\beta$, meaning that even though it is
a one-loop effect, it can be important at large $\tan\beta$.

The radiatively-corrected couplings can be parameterized by an effective
Lagrangian~\cite{Carena:2001bg},
\begin{equation}
  - \mathcal{L}_{\rm eff} = \epsilon_{ij} h_b \bar b_R H_d^i Q_L^j
  + \Delta h_b \bar b_R Q_L^k H_u^{k *} + {\rm h.c.},
\end{equation}
with $H_u$ and $H_d$ defined in the usual way for the MSSM with
opposite hypercharges.  Note that we absorb into $h_b$ any SUSY
radiative corrections to the ``right-Higgs'' coupling.

The physical bottom quark mass is given by
\begin{equation}
  m_b = \frac{h_b v_d}{\sqrt{2}} + \frac{\Delta h_b v_u}{\sqrt{2}}
  = \frac{h_b v_{SM} \cos\beta}{\sqrt{2}}
  \left( 1 + \frac{\Delta h_b \tan\beta}{h_b} \right)
  \equiv \frac{h_b v_{SM} \cos\beta}{\sqrt{2}} \left( 1 + \Delta_b \right).
\end{equation}
(Note the factor of $\tan\beta$ that is absorbed into the definition
of $\Delta_b$.)  Similarly, the $h^0 b \bar b$ coupling becomes
\begin{equation}
  g_b^h = - \sin\alpha \frac{h_b}{\sqrt{2}}
  + \cos\alpha \frac{\Delta h_b}{\sqrt{2}},
\end{equation}
where the mixing angle $\alpha$ is defined as in
Eq.~\ref{eq:2HDM2alphadefinition} for the 2HDM-II.  This coupling can
be written in terms of $m_b$ and $\Delta_b$ by noting that
\begin{equation}
  \frac{h_b}{\sqrt{2}} = \frac{m_b}{v_{SM} \cos\beta} \frac{1}{1 + \Delta_b},
  \qquad \qquad
  \frac{\Delta h_b}{\sqrt{2}} = \frac{m_b}{v_{SM} \sin\beta}
  \frac{\Delta_b}{1 + \Delta_b}.
\end{equation}
Inserting these relations into $g_b^h$ and normalizing by the SM
coupling we obtain
\begin{equation}
  \bar g_b = \sin(\beta-\alpha)
  - \tan\beta \cos(\beta-\alpha) 
  \frac{1 - \cot^2\beta \, \Delta_b}{1 + \Delta_b}.
\end{equation}

The $\Delta_b$ corrections are typically the only large SUSY radiative
corrections to the Higgs Yukawa
couplings~\cite{Carena:1998gk,Carena:2001bg} -- in particular, the
analogous corrections to the Higgs couplings to top quarks are not
$\tan\beta$ enhanced, and those to the Higgs couplings to tau leptons
involve only the small electroweak gauge
couplings~\cite{Pierce:1996zz}.  Thus the SUSY corrections to these
couplings can generally be neglected, and the usual 2HDM-II relations
are recovered:
\begin{eqnarray}
\bar g_W &=& \sin(\beta-\alpha),
\nonumber \\
\bar g_u &=& \frac{\cos\alpha}{\sin\beta}
= \sin(\beta-\alpha) + \cot\beta \cos(\beta-\alpha),
\nonumber \\
\bar g_{\ell} &=& -\frac{\sin\alpha}{\cos\beta}
= \sin(\beta-\alpha) - \tan\beta \cos(\beta - \alpha).
\end{eqnarray}

The model parameters can be obtained as in the 2HDM-II by using the
couplings that are unaffected by $\Delta_b$:
\begin{equation}
\tan\beta = \left[
  \frac{\bar g_{\ell}^2 - 1}{1 - \bar g_u^2} \right]^{1/2},
  \qquad
\cos\alpha = \bar g_u \left[
  \frac{1 - \bar g_{\ell}^2}{\bar g_u^2 - \bar g_{\ell}^2} \right]^{1/2},
  \qquad
  \sin\alpha = \bar g_{\ell} \left[
    \frac{1 - \bar g_u^2}{\bar g_{\ell}^2 - \bar g_u^2} \right]^{1/2}.
\end{equation}
The value of $\Delta_b$ can also be extracted from the $h$ couplings
using~\cite{Carena:2001bg}
\begin{equation}
  \Delta_b = \frac{\bar g_b - \bar g_{\ell}}{\bar g_u - \bar g_b}.
\end{equation}

We first note that the pattern relation of the 2HDM-II among the
couplings $\bar g_W$, $\bar g_u$ and $\bar g_{\ell}$ survives:
\begin{equation}
  P_{u\ell} \equiv
  \bar g_W (\bar g_u + \bar g_{\ell}) - \bar g_u \bar g_{\ell} = 1.
  \label{eq:MSSMtaupattrel}
\end{equation}
This allows the MSSM Higgs sector to be distinguished from the more
general three-Higgs-doublet models discussed in the next section and
allows one to test for the presence of additional singlets or doublets
that mix with $h$ or carry nonzero vevs.

However, the $\Delta_b$ correction to the Higgs coupling to bottom
quarks leads to $\bar g_d \neq \bar g_{\ell}$, such that the pattern
relation among the $W$, $u$ and $d$ couplings is violated:
\begin{equation}
P_{ud} \equiv \bar g_W (\bar g_u + \bar g_d) - \bar g_u \bar g_d
= 1 - \cos^2(\beta-\alpha) \frac{\Delta_b (1 + \cot^2\beta)}{1 + \Delta_b}.
\label{eq:MSSMDbpattrel}
\end{equation}
Depending on the sign of $\Delta_b$, the right-hand side can be
greater or less than one. 

In the decoupling limit the deviations of the $h$ couplings from their
SM values can be parameterized in terms of $\delta \equiv \cos(\beta -
\alpha)$.  We have,
\begin{eqnarray}
&\bar g_W = \sqrt{1 - \delta^2} \simeq 1 - \frac{1}{2} \delta^2,
\qquad \qquad
&\frac{\Gamma_W^h}{\Gamma_W^{SM}} = 1 - \delta^2, \nonumber \\
&\bar g_u = \sqrt{1 - \delta^2} + \cot\beta \, \delta
\simeq 1 + \cot\beta \, \delta,
\qquad \qquad
&\frac{\Gamma_u^h}{\Gamma_u^{SM}} \simeq 1 + 2 \cot\beta \, \delta,
\nonumber \\
&\bar g_b = \sqrt{1 - \delta^2} - \tan\beta^{\prime} \, \delta
\simeq 1 - \tan\beta^{\prime} \, \delta,
\qquad \qquad
&\frac{\Gamma_b^h}{\Gamma_b^{SM}} \simeq 1 - 2 \tan\beta^{\prime} \, \delta,
\nonumber \\
&\bar g_{\ell} = \sqrt{1 - \delta^2} - \tan\beta \, \delta
\simeq 1 - \tan\beta \, \delta,
\qquad \qquad
&\frac{\Gamma_{\ell}^h}{\Gamma_{\ell}^{SM}} \simeq 1 - 2 \tan\beta \, \delta.
\end{eqnarray}
Note that $\delta$ can take either sign.  These expressions are
identical to those given in Eq.~\ref{eq:2HDMIIdecoup} for the 2HDM-II
except that in the bottom quark couplings we have replaced $\tan\beta$
with
\begin{equation}
\tan\beta^{\prime} \equiv \tan\beta \,
\frac{1 - \cot^2\beta \, \Delta_b}{1 + \Delta_b}.
\end{equation}

\section{Fermion masses from three doublets}
\label{sect:3hdm}

Finally we consider a model in which the fermion masses arise
``democratically'' from couplings to three different Higgs doublets --
i.e., models in which the masses of the up-type quarks, down-type
quarks, and charged leptons are generated by couplings to three
different Higgs doublets $\Phi_u$, $\Phi_d$, and $\Phi_{\ell}$,
respectively.  Such a model was considered in
Ref.~\cite{Grossman:1994jb}.  We also consider extensions of this
model obtained by adding one or more singlets or doublets that do not
couple to fermions.\footnote{Models with three or more doublets
introduce the possibility of new CP-violating parameters in the
$n\times n$, $n\geq 3$ mixing matrices of the charged
scalars~\cite{Grossman:1994jb}.  Again, we neglect the
possibility of CP-violating effects in this work.}

A similar Higgs-fermion coupling structure has recently been proposed
in the ``Private Higgs'' model~\cite{Porto:2007ed,Porto:2008hb}, which
introduces one Higgs doublet for each of the six flavors of quarks in
order to address the hierarchy of quark masses.  Here, however, we
limit the discussion to models in which the masses of fermions of a
given electric charge are generated by their couplings to only one
Higgs doublet; in particular, we do not allow the Higgs coupling
structure to differ by fermion generation.  We also make no
assumptions about the structure of the Higgs potential.  The
discussion here can be extended to models in which the three
generations are treated differently but care must be taken to avoid
Higgs-mediated flavor-changing neutral currents.

\subsection{Democratic Three Higgs Doublet Model (3HDM-D)}

In this model the up-type quarks, down-type quarks, and charged
leptons each get their mass from a different Higgs doublet, denoted
$\Phi_u$, $\Phi_d$, and $\Phi_{\ell}$, respectively.  The constraints
of Eqs.~\ref{eq:sasum}, \ref{eq:sbsum} become
\begin{equation}
  a_u^2 + a_d^2 + a_{\ell}^2 = 1, \qquad \qquad
  b_u^2 + b_d^2 + b_{\ell}^2 = 1.
\end{equation}
The normalized couplings of $h$ to SM particles are then given by
\begin{equation}
  \bar g_W = a_u b_u + a_d b_d + a_{\ell} b_{\ell}, \qquad \qquad
  \bar g_u = \frac{a_u}{b_u}, \qquad \qquad
  \bar g_d = \frac{a_d}{b_d}, \qquad \qquad
  \bar g_{\ell} = \frac{a_{\ell}}{b_{\ell}}.
  \label{eq:3HDMcoups}
\end{equation}
The vev ratios $b_u$, $b_d$, and $b_{\ell}$ can all be chosen real and
positive.  Simultaneously, we can choose $\bar g_W$ positive through
an appropriate rephasing of the mass eigenstate $h$.  
There is no freedom left to choose the signs of the fermion couplings
$\bar g_u$, $\bar g_d$, or $\bar g_{\ell}$; depending on the underlying
values of the parameters they can take any combination of signs so long
as at least one of them is positive.

The constraint equations and coupling relations can be solved explicitly
for the $b_i$ factors in terms of the $h$ couplings:
\begin{eqnarray}
b_u &=& \left[ \frac{1 - \bar g_W (\bar g_d + \bar g_{\ell})
    + \bar g_d \bar g_{\ell}}
  {(\bar g_u - \bar g_d)(\bar g_u - \bar g_{\ell})} \right]^{1/2},
  \nonumber \\
b_d &=& \left[ \frac{1 - \bar g_W (\bar g_u + \bar g_{\ell})
    + \bar g_u \bar g_{\ell}}
  {(\bar g_d - \bar g_u)(\bar g_d - \bar g_{\ell})} \right]^{1/2},
  \nonumber \\
b_{\ell} &=& \left[ \frac{1 - \bar g_W (\bar g_u + \bar g_d)
    + \bar g_u \bar g_d}
  {(\bar g_{\ell} - \bar g_u)(\bar g_{\ell} - \bar g_d)} \right]^{1/2}.
\label{eq:3HDMbi}
\end{eqnarray}
We also obtain solutions for the $a_i$ factors,
\begin{equation}
a_u = b_u \bar g_u, \qquad \qquad
a_d = b_d \bar g_d, \qquad \qquad
a_{\ell} = b_{\ell} \bar g_{\ell}.
\end{equation}
As in the other solvable models, if the relative signs of $\bar g_u$,
$\bar g_d$, $\bar g_{\ell}$ and $\bar g_W$ are known, then the
solution for the model parameters is unique.  However, if the relative
signs are not known, discrete ambiguities arise in the solutions
for the $b_i$ and $a_i$ factors.

The key feature that distinguishes the democratic 3HDM from the
previous models considered is that $\bar g_u \neq \bar g_d \neq \bar
g_{\ell}$.  While this is also true in the MSSM with $\Delta_b$
corrections, the MSSM couplings satisfy the pattern relation
$P_{u\ell} = 1$ involving the couplings $\bar g_W$, $\bar g_u$, and $\bar
g_{\ell}$ (Eq.~\ref{eq:MSSMtaupattrel}); this relation does not hold
in the 3HDM-D.

The features of this model can be clarified by examining its parallels
with the 2HDM-II+D.  In particular, the behavior of $\bar g_W$, $\bar
g_u$ and $\bar g_d$ is identical to that in the 2HDM-II+D, while now
$\bar g_{\ell}$ behaves differently with the model parameters and
serves to distinguish the current model.  As in the 2HDM-II+D, we
parameterize the mixing according to $h = \cos\theta \, h^{\prime} +
\sin\theta \, \phi_{\ell}$, with $h^{\prime} \equiv \cos\alpha \,
\phi_u - \sin\alpha \, \phi_d$, yielding $a_u = \cos\theta
\cos\alpha$, $a_d = -\cos\theta \sin\alpha$, and $a_{\ell} =
\sin\theta$.  We also define $\tan\beta \equiv v_u/v_d = b_u/b_d$ and
$\cos\Omega \equiv \sqrt{b_u^2 + b_d^2}$, $\sin\Omega = b_{\ell}$.  The
couplings of $h$ are then given by
\begin{eqnarray}
  \bar g_W &=& \cos\Omega \cos\theta \sin(\beta-\alpha)
  + \sin\Omega \sin\theta, \nonumber \\
  \bar g_u &=& \frac{\cos\theta}{\cos\Omega} \frac{\cos\alpha}{\sin\beta},
  \qquad \qquad
  \bar g_d = -\frac{\cos\theta}{\cos\Omega} \frac{\sin\alpha}{\cos\beta},
  \qquad \qquad
  \bar g_{\ell} = \frac{\sin\theta}{\sin\Omega}.
\end{eqnarray}

The decoupling relations can be parameterized exactly as in the 2HDM-II+D
(Eq.~\ref{eq:decoup2HDM2D}) except for $\bar g_{\ell}$, which is given by
\begin{equation}
  \bar g_{\ell} = \frac{\langle h | \phi_{\ell} \rangle}{b_{\ell}}
  = \sqrt{1-\delta^2} 
  + \delta \left[ \cos\gamma \cot\Omega \right],
\end{equation}
where $\phi_{\ell}$ is the third doublet and $\gamma$ is defined as in
Eq.~\ref{eq:gammadef} with $\phi_0 \to \phi_{\ell}$.

\subsection{3HDM-D plus one or more singlets (3HDM-D+S)}

We now consider the consequences of adding a real singlet scalar
field, $S$, to the 3HDM-D.  The constraints of Eqs.~\ref{eq:sasum},
\ref{eq:sbsum} become
\begin{equation}
  a_u^2 + a_d^2 + a_{\ell}^2 + a_s^2 = 1, \qquad \qquad
  b_u^2 + b_d^2 + b_{\ell}^2 = 1.
\end{equation}
The formulae for the normalized
couplings of $h$ to SM particles are identical to those of the 3HDM-D
given in Eq.~\ref{eq:3HDMcoups}.
The vev ratios $b_u$, $b_d$ and $b_{\ell}$ and the coupling $\bar g_W$
can all be chosen real and positive; $a_s$ can then be chosen real and 
positive by an appropriate rephasing of $S$.

Because of the presence of the additional parameter $a_s$, this model
is distinguishable from the 3HDM-D in part of its parameter space, as
we now show.  First we define the following three combinations of $h$
couplings,
\begin{eqnarray}
X_u &\equiv& \frac{1 - \bar g_W (\bar g_d + \bar g_{\ell})
    + \bar g_d \bar g_{\ell}}
  {(\bar g_u - \bar g_d)(\bar g_u - \bar g_{\ell})}
  = b_u^2 + \frac{a_s^2}{(\bar g_u - \bar g_d)(\bar g_u - \bar g_{\ell})},
  \nonumber \\
X_d &\equiv& \frac{1 - \bar g_W (\bar g_u + \bar g_{\ell})
    + \bar g_u \bar g_{\ell}}
  {(\bar g_d - \bar g_u)(\bar g_d - \bar g_{\ell})}
  = b_d^2 + \frac{a_s^2}{(\bar g_d - \bar g_u)(\bar g_d - \bar g_{\ell})},
  \nonumber \\
X_{\ell} &\equiv& \frac{1 - \bar g_W (\bar g_u + \bar g_d)
    + \bar g_u \bar g_d}
  {(\bar g_{\ell} - \bar g_u)(\bar g_{\ell} - \bar g_d)}
  = b_{\ell}^2
  + \frac{a_s^2}{(\bar g_{\ell} - \bar g_u)(\bar g_{\ell} - \bar g_d)}.
\label{eq:Xi}
\end{eqnarray}
Here $X_u + X_d + X_{\ell} = 1$ by construction.  Note that if these
formulae were applied to the 3HDM-D, they would yield $b_u^2$, $b_d^2$
and $b_{\ell}^2$, respectively (cf.~Eq.~\ref{eq:3HDMbi}; in the
current model this is recovered in the limit $a_s \to 0$).  In
particular, the values of all three $X_i$ would necessarily lie
between zero and one.  However, in part of the parameter space of the
3HDM-D+S, one of the $X_i$ can be negative.  In this part of the
parameter space, if one were to (incorrectly) assume the 3HDM-D and
attempt to solve for the $b_i$, Eq.~\ref{eq:3HDMbi} would fail to
yield a solution.  Thus we see that the footprint of the 3HDM-D+S in
the space of $h$ couplings is larger than that of the 3HDM-D, and
therefore the model with an additional singlet can be distinguished
from the 3HDM-D in part of its parameter space.

A negative value for one of the $X_i$ can occur because exactly one
of the three denominators in Eq.~\ref{eq:Xi} is negative.  This allows
us to obtain a lower limit on $a_s^2$ when one of the $X_i$ is negative.
We first define
\begin{equation}
Y = \left\{ \begin{array}{cc}
  (\bar g_u - \bar g_d)(\bar g_u - \bar g_{\ell}) X_u \qquad &
  {\rm if} \ X_u < 0, \\
  (\bar g_d - \bar g_u)(\bar g_d - \bar g_{\ell}) X_d \qquad &
  {\rm if} \ X_d < 0, \\
  (\bar g_{\ell} - \bar g_u)(\bar g_{\ell} - \bar g_d) X_{\ell} \qquad &
  {\rm if} \ X_{\ell} < 0.
  \end{array} \right.
\label{eq:Y}
\end{equation}
Note $0 < Y < 1$ by construction for parameter points where $Y$ is
defined.  These expressions are entirely determined in terms of the
$h$ couplings.  The lower limit on $a_s^2$ is then given by
$a_s^2 \geq Y$.

For completeness, we give here the relations for the parameters $b_i$
and $a_i$ in terms of the $h$ couplings and $\xi \equiv 1 - a_s^2$:
\begin{eqnarray}
b_u &=& \left[ \frac{\xi - \bar g_W (\bar g_d + \bar g_{\ell})
    + \bar g_d \bar g_{\ell}}
  {(\bar g_u - \bar g_d)(\bar g_u - \bar g_{\ell})} \right]^{1/2},
  \nonumber \\
b_d &=& \left[ \frac{\xi - \bar g_W (\bar g_u + \bar g_{\ell})
    + \bar g_u \bar g_{\ell}}
  {(\bar g_d - \bar g_u)(\bar g_d - \bar g_{\ell})} \right]^{1/2},
  \nonumber \\
b_{\ell} &=& \left[ \frac{\xi - \bar g_W (\bar g_u + \bar g_d)
    + \bar g_u \bar g_d}
  {(\bar g_{\ell} - \bar g_u)(\bar g_{\ell} - \bar g_d)} \right]^{1/2},
\nonumber \\
a_u &=& b_u \bar g_u, \qquad \qquad
a_d = b_d \bar g_d, \qquad \qquad
a_{\ell} = b_{\ell} \bar g_{\ell}.
\end{eqnarray}
Because $a_s$ cannot be uniquely determined in this model, the model
is underconstrained and the parameters $b_i$ and $a_i$ cannot be
uniquely extracted.

These results can easily be extended to models containing two or more
singlets by making the replacement
\begin{equation}
  a_s^2 \rightarrow \sum_{\rm singlets} a_{s_i}^2 = 1 - \xi.
\end{equation}
We see that it is not possible to tell whether only one singlet or
more than one singlet has been added to the 3HDM-D on the basis of $h$
couplings alone.

\subsection{3HDM-D plus additional doublet(s)}

We finally consider the consequences of adding an additional Higgs
doublet $\Phi_0$ to the 3HDM-D.  The additional doublet carries a vev
but does not couple to fermions.  The constraint equations become,
\begin{equation}
  a_u^2 + a_d^2 + a_{\ell}^2 + a_0^2 = 1, \qquad \qquad
  b_u^2 + b_d^2 + b_{\ell}^2 + b_0^2 = 1.
\end{equation}
The normalized couplings of $h$ to SM particles are given by
\begin{equation}
  \bar g_W = a_u b_u + a_d b_d + a_{\ell} b_{\ell} + a_0 b_0, \qquad \qquad
  \bar g_u = \frac{a_u}{b_u}, \qquad \qquad
  \bar g_d = \frac{a_d}{b_d}, \qquad \qquad
  \bar g_{\ell} = \frac{a_{\ell}}{b_{\ell}}.
\end{equation}
All four $b_i$ parameters and $\bar g_W$ can be chosen real and
positive, while now $\bar g_u$, $\bar g_d$, and $\bar g_{\ell}$ can
have any combination of signs; in particular, all three of these
couplings can be negative if $a_0 b_0$ is big enough to keep $\bar
g_W$ positive.

Like the 3HDM-D+S, this model is distinguishable from the 3HDM-D in
part of its parameter space; the parameters and couplings of the
current model reduce to the form of the 3HDM-D+S in the limit $b_0 \to
0$.  However, the footprint of the current model in $h$ coupling space
is larger than that of the 3HDM-D+S, so that in part of the parameter
space the presence of the extra doublet can be detected, as we now show.

We again define $X_u$, $X_d$ and $X_{\ell}$ in terms of the $h$
couplings as in Eq.~\ref{eq:Xi}.  In terms of the underlying model
parameters, these can be expressed as
\begin{eqnarray}
X_u &=& b_u^2 + \frac{a_0^2 + b_0^2 \bar g_d \bar g_{\ell}
  - a_0 b_0 (\bar g_d + \bar g_{\ell})}
{(\bar g_u - \bar g_d)(\bar g_u - \bar g_{\ell})}, \nonumber \\
X_d &=& b_d^2 + \frac{a_0^2 + b_0^2 \bar g_u \bar g_{\ell}
  - a_0 b_0 (\bar g_u + \bar g_{\ell})}
{(\bar g_d - \bar g_u)(\bar g_d - \bar g_{\ell})}, \nonumber \\
X_{\ell} &=& b_{\ell}^2 + \frac{a_0^2 + b_0^2 \bar g_u \bar g_{\ell}
  - a_0 b_0 (\bar g_u + \bar g_d)}
{(\bar g_{\ell} - \bar g_u)(\bar g_{\ell} - \bar g_d)}.
\end{eqnarray}
Again, $X_u + X_d + X_{\ell} = 1$ by construction.  In the limit $b_0
\to 0$, these expressions reduce to those for the 3HDM-D+S given in
Eq.~\ref{eq:Xi}; in that limit the numerator of the second term is
just $a_0^2$, which must lie between zero and one.  When $b_0 \neq 0$,
however, the numerator of the second term can be less than zero or
greater than one. 

In the part of parameter space with one negative $X_i$ we again
construct the quantity $Y$ as given in Eq.~\ref{eq:Y}.  In the
3HDM-D+S, $Y$ provided a lower bound for $a_0^2$; in particular $0
\leq Y \leq 1$ always.  In the current model, however, one can also
obtain $Y < 0$ (when $X_i$ is negative due to the numerator of the
second term being negative) or $Y > 1$ (when $X_i$ is negative due to
the denominator of the second term being negative and the numerator of
the second term is sufficiently greater than one).  Neither of these
possibilities can occur in the 3HDM-D+S and they therefore allow the
current model to be distinguished in part of its parameter space.

The analysis can easily be extended to the 3HDM-D plus two or more
doublets.  We have seen in the case of the 2HDM-I plus an additional
doublet (Sec.~\ref{sec:2HDM-I+D}) and the 2HDM-II plus additional
doublets (Sec.~\ref{sec:2HDM-II+D}) that, once the model already
contains one doublet that does not couple to fermions, adding
additional doublets that do not couple to fermions does not change the
model footprint in $h$ coupling space.  The same is true for the
3HDM-D plus additional doublets.  Adding one additional doublet
changes the model footprint as we have seen.  Adding a second
additional doublet, however, does not further change the model
footprint; thus it is not possible to tell how many additional
doublets have been added to the 3HDM-D based only on the $h$
couplings.

The addition of singlet(s) to the 3HDM-D plus a doublet can be dealt
with in a similar way.  As far as the couplings of $h$ are concerned,
adding a singlet is indistinguishable from adding an additional Higgs
doublet with zero vev.  We see then that it is not possible to tell
whether singlets have been added to the 3HDM-D plus a
doublet based only on the $h$ couplings.

\section{Radiative Corrections}
\label{sect:radcorr}

In order to translate between the tree-level Lagrangian parameters
$\bar g_W$, $\bar g_u$, $\bar g_d$, and $\bar g_{\ell}$ and
experimentally observable $h$ production cross sections and decay
partial widths, radiative corrections must be included.  This program
has been carried out in great detail for the SM Higgs as well as for
the MSSM.  For more general multi-Higgs-doublet models, however,
detailed results are lacking; such corrections would be needed for a
translation between observables and the underlying Lagrangian
parameters at the few-percent level.  We can however make the
following general observations.  

QCD corrections are universal and can be taken over from the SM,
assuming that no new strongly-interacting particles contribute.  In
the MSSM, for example, squarks and gluinos yield large flavor-specific
radiative corrections; integrating out these contributions into an
effective Lagrangian yields the $\Delta_b$ formalism but leads to a
violation of the underlying natural flavor conservation of the MSSM
Higgs sector.

Electroweak radiative corrections are not universal and in principle
must be computed for each model.  These depend on the model content --
both the extended Higgs sector and any additional new physics that may
be present.  Some parts of these corrections can be simply absorbed
into our parameterization; for example, the largest electroweak
corrections to the MSSM Higgs sector from top quark and top squark
loops can be absorbed into an effective Higgs sector mixing angle
$\alpha$.  However, vertex corrections remain an issue for precision
parameter extraction.

Experimental determination of the relative signs of $h$ couplings is
also potentially problematic.  These signs are accessible only through
the interference of competing amplitudes in loops, and thus
nonstandard sign combinations can be masked or faked by additional new
contributions to loop-induced couplings.

Throughout we choose the phase of $h$ such that $\bar g_W$ is
positive.  The sign of $\bar g_u$ is then accessible through the
$h\gamma\gamma$ coupling: in the SM, the $W$ loop dominates while the
top quark loop interferes destructively, reducing the $h \to
\gamma\gamma$ partial width by $\sim 30\%$.  The relative sign of
$\bar g_d$ and $\bar g_u$ is in principle accessible from the $ggh$
coupling: again in the SM the top quark loop dominates, while
top-bottom interference is about a 10\% effect for moderate Higgs
masses; however, the QCD scale uncertainty is still of this order.
The sign of $\bar g_{\ell}$ will be even more difficult, since its
contribution to $h\gamma\gamma$ is extremely small.  The loop-induced
$h \gamma Z$ coupling would provide additional information, but its
experimental detection does not seem feasible at the moment.  The best
strategy may be to examine all possible sign combinations for $h$
couplings and enumerate their implications for the underlying model
parameters and the size of the possible new physics contributions to
loop-induced $h$ couplings.

\section{Discussion and conclusions}
\label{sect:disc}

Our ultimate aim in this work is to provide a framework for
distinguishing among competing models for the Higgs sector.  To this
end we have studied the patterns of tree-level couplings of a single
CP-even state $h$ in all models that can be constructed out of
SU(2)$_L$ doublets and/or singlets, subject to the requirement of
natural flavor conservation.  Distinguishing one model from another
relies not only on the underlying theoretical distinctions between the
values taken by the $h$ couplings, but also on the experimental and
theoretical precision with which those couplings can be measured.
Here we collect our theoretical results, then turn to the question of
model discrimination based on experimental data.

Our theoretical results can be conveniently summarized in the form of
a decision tree, as follows.  We assume a deviation from the SM; $n
\geq 1$ counts additional singlets (S) and $m \geq 1$ counts
additional doublets (D) that do not couple to fermions.  We denote the
pattern relation involving fermion couplings $\bar g_i$ and $\bar g_j$
as $P_{ij} \equiv \bar g_W (\bar g_i + \bar g_j) - \bar g_i \bar g_j$.
The $X_i$ factors are defined in Eq.~\ref{eq:Xi} and $Y$ is defined in
Eq.~\ref{eq:Y}.
\begin{enumerate}
\item $\bar g_u = \bar g_d = \bar g_{\ell}$ (Type-I--like)
\begin{enumerate}
  \item $\bar g_W = \bar g_f$: SM+$n$S; 2HDM-I when $\langle
    \Phi_0 \rangle = 0$.
  \item $\bar g_W \neq \bar g_f$: 2HDM-I; 2HDM-I+$n$S; 2HDM-I+$m$D;
    2HDM-I+$n$S+$m$D.
\end{enumerate}
\item $\bar g_d = \bar g_{\ell} \neq \bar g_u$ (Type-II--like)
\begin{enumerate}
  \item $P_{ud} = P_{u \ell} = 1$: 2HDM-II.
  \item $0 \leq P_{ud} = P_{u \ell} \leq 1$: 2HDM-II+$n$S;
  2HDM-II+$m$D; 2HDM-II+$n$S+$m$D.
  \item $P_{ud} = P_{u \ell} > 1$ or $P_{ud} = P_{u \ell} < 0$:
  2HDM-II+$m$D; 2HDM-II+$m$D+$n$S.
\end{enumerate}
\item $\bar g_u = \bar g_{\ell} \neq \bar g_d$ (flipped 2HDM--like)
\begin{enumerate}
  \item $P_{ud} = P_{\ell d} = 1$: flipped 2HDM.
  \item $0 \leq P_{ud} = P_{\ell d} \leq 1$: flipped
  2HDM+$n$S; flipped 2HDM+$m$D; flipped
  2HDM+$n$S+$m$D.
  \item $P_{ud} = P_{\ell d} > 1$ or $P_{ud} = P_{\ell d} < 0$:
  flipped 2HDM+$m$D; flipped 2HDM+$m$D+$n$S.
\end{enumerate}
\item $\bar g_u = \bar g_d \neq \bar g_{\ell}$ (lepton-specific 2HDM--like)
\begin{enumerate}
  \item $P_{u \ell} = P_{d \ell} = 1$: lepton-specific 2HDM.
  \item $0 \leq P_{u \ell} = P_{d \ell} \leq 1$: 
  lepton-specific 2HDM+$n$S; 
  lepton-specific 2HDM+$m$D; lepton-specific 2HDM+$n$S+$m$D.
  \item $P_{u \ell} = P_{d \ell} > 1$ or $P_{u \ell} = P_{d \ell} <
  0$: lepton-specific 2HDM+$m$D;
  lepton-specific 2HDM+$m$D+$n$S.
\end{enumerate}
\item $\bar g_u \neq \bar g_d \neq \bar g_{\ell}$
\begin{enumerate}
  \item $P_{u \ell} = 1$: MSSM with $\Delta_b$.
  \item $P_{u \ell} \neq 1$
    \begin{enumerate}
    \item $0 \leq X_i \leq 1$: 3HDM-D; 3HDM-D+$n$S; 3HDM-D+$m$D;
    3HDM-D+$n$S+$m$D.
    \item One of $X_i < 0$ and $0 \leq Y \leq 1$: 3HDM-D+$n$S;
    3HDM-D+$m$D; 3HDM-D+$n$S+$m$D.
    \item One of $X_i < 0$ and $Y < 0$ or $Y > 1$: 3HDM-D+$m$D;
    3HDM-D+$m$D+$n$S.
    \end{enumerate}
\end{enumerate}
\end{enumerate}
In particular, we count 15 models (or sets of models) that are
distinguishable in principle based on the couplings of $h$.
Explicit formulae for $h$ partial widths (equivalently couplings
squared) in the decoupling limit, i.e., for small deviations from the
SM predictions, are collected in Table~\ref{tab:expansions}.

\begin{table}[htdp]
\begin{center}
\begin{tabular}{|c|cccc|}
\hline \hline
Model &  $\Gamma_W^h/\Gamma_W^{SM}$ & $\Gamma_d^h/\Gamma_d^{SM}$ 
  & $\Gamma_u^h/\Gamma_u^{SM}$ & $\Gamma_{\ell}^h/\Gamma_{\ell}^{SM}$ \\
\hline
SM & 1&1&1&1\\
SM+S & $1-\delta^2$&$1-\delta^2$&$1-\delta^2$&$1-\delta^2$\\
2HDM-I & $1-\delta^2$&$1+2\delta/t_\beta$&$1+2\delta/t_\beta$&$1+2\delta/t_\beta$\\
2HDM-II & $1-\delta^2$&$1-2t_\beta\delta$&$1+2\delta/t_\beta$&$1-2t_\beta\delta$\\
2HDM-II+S & $1-\delta^2-\epsilon^2$&$1-2t_\beta\delta-\epsilon^2$&$1+2\delta/t_\beta-\epsilon^2$&$1-2t_\beta\delta-\epsilon^2$\\
2HDM-II+D 
  & $1 - \delta^2$ 
  & $1 - 2 \delta (s_{\gamma} t_{\beta}/c_{\Omega} + c_{\gamma} t_{\Omega})$ 
  & $1 + 2 \delta (s_{\gamma}/c_{\Omega} t_{\beta} - c_{\gamma} t_{\Omega})$
  & $1 - 2 \delta (s_{\gamma} t_{\beta}/c_{\Omega} + c_{\gamma} t_{\Omega})$ \\
Flipped 2HDM &  $1-\delta^2$&$1-2t_\beta\delta$&$1+2\delta/t_\beta$&$1+2\delta/t_\beta$\\
Lepton-specific 2HDM & $1-\delta^2$&$1+2\delta /t_\beta$ & $1+2\delta /t_\beta$ & $1-2t_\beta\delta$\\
MSSM & $1-\delta^2$ & $1-2t_{\beta}^{\prime} \delta$ & $1+2\delta/t_{\beta}$ & $1-2t_{\beta}\delta$\\
3HDM-D 
  & $1 - \delta^2$ 
  & $1 - 2 \delta (s_{\gamma} t_{\beta}/c_{\Omega} + c_{\gamma} t_{\Omega})$ 
  & $1 + 2 \delta (s_{\gamma}/c_{\Omega} t_{\beta} - c_{\gamma} t_{\Omega})$
  & $1 + 2 \delta c_{\gamma}/t_{\Omega}$ \\
\hline \hline
\end{tabular}
\caption{\label{tab:expansions}
Behavior of the Higgs partial widths (equivalently couplings
squared) near the decoupling limit, $|\delta| \ll 1$.  For the
2HDM-II+S we also require $\epsilon^2 \ll 1$.  The other parameters
are defined as $t_{\beta} \equiv \tan\beta = v_f/v_0$ in the 2HDM-I,
$v_u/v_d$ in the 2HDM-II, flipped 2HDM, and 3HDM-D, and $v_q/v_{\ell}$
in the lepton-specific 2HDM.  For the MSSM we define
$t_{\beta}^{\prime} \equiv \tan\beta^{\prime} \equiv v_u (1-\cot^2
\beta \, \Delta_b ) / v_d(1+\Delta_b)$.  For the 2HDM-II+D and 3HDM-D
we also define $c_{\Omega} \equiv \cos\Omega = \sqrt{v_u^2 +
v_d^2}/v_{SM}$ and $\gamma$ is the remaining mixing angle that
parameterizes the state $h$.}
\end{center}
\end{table}

\begin{figure}[htbp]
\begin{center}
\includegraphics[width=0.8\textwidth,angle=0]{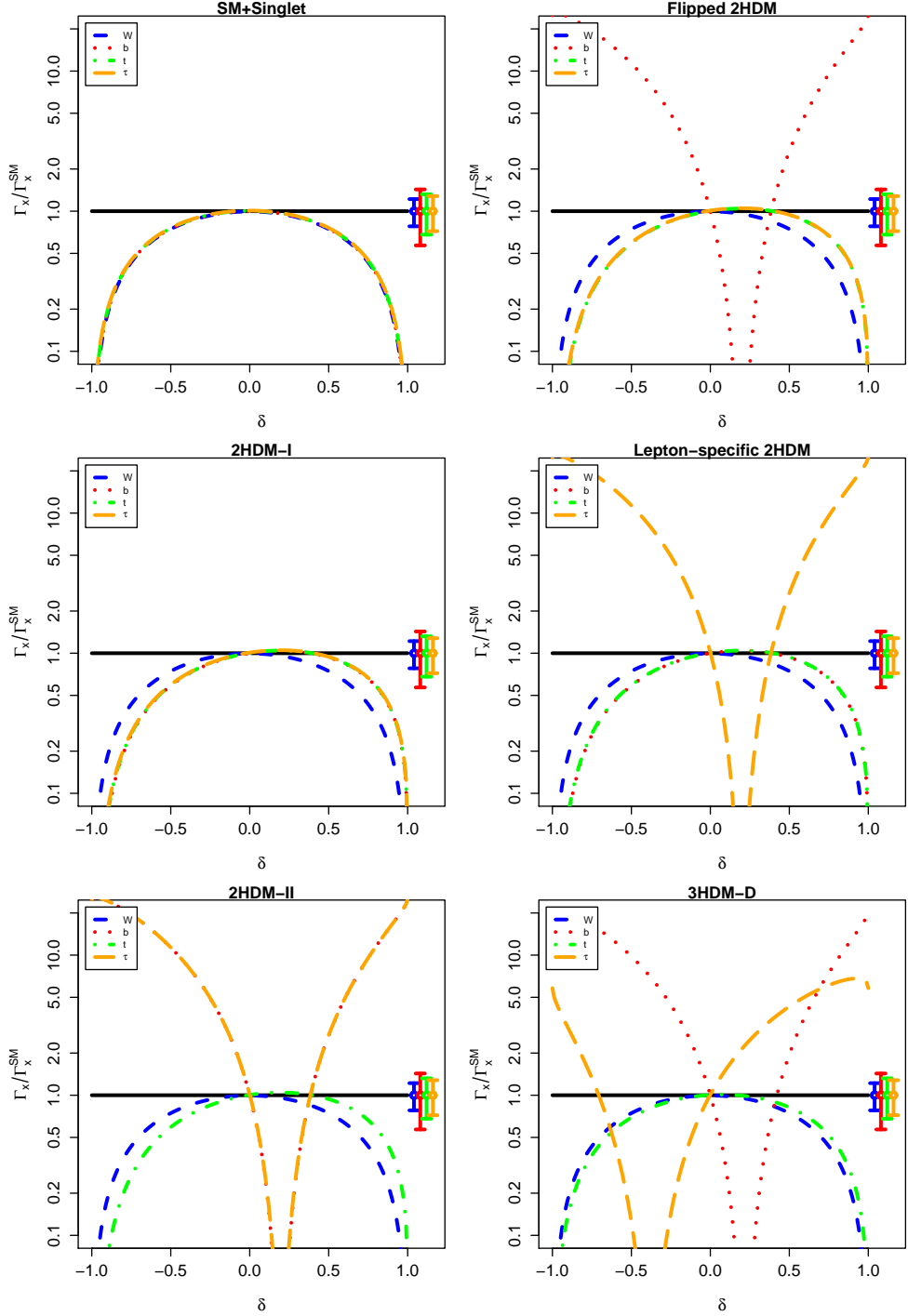}
\caption{Higgs partial widths (equivalently couplings squared) as a
function of the parameter $\delta$ in the SM+singlet, the Type-I,
Type-II, flipped and lepton-specific 2HDMs, and the democratic 3HDM.
We set $\tan\beta = 5$ for all models except the SM+S; for the 3HDM-D
we also set $\sin\Omega = 0.2$ and $\cos\gamma=0.5$.
At the right of each panel we show the expected $1\sigma$ LHC
measurement uncertainties for the Higgs couplings-squared to $WW$
(blue), $bb$ (red), $tt$ (green), and $\tau\tau$ (orange) for
$m_h=120$ GeV and SM event rates, taken from
Table~\ref{tab:exptuncerts}.  Note the log scale on the $y$ axis.}
\label{fig:hmodels-delta}
\end{center}
\end{figure}

We now give a first comparison of our theoretical results to
anticipated LHC and ILC measurements of Higgs couplings-squared.  In
Fig.~\ref{fig:hmodels-delta} we illustrate the behavior of the partial
widths (equivalently couplings squared) of $h$ to $W$ or $Z$ boson
pairs, down-type quarks, up-type quarks, and charged leptons,
normalized to their SM values, as a function of the decoupling
parameter $\delta$.  Because we consider the full range $-1 < \delta <
1$, we use the exact formulae from the text rather than the decoupling
limit approximations of Table~\ref{tab:expansions}.  We show results
for the SM plus a singlet (Eq.~\ref{eq:SM+Sdecoup}), the Type-I 2HDM
(Eqs.~\ref{eq:2HDMIdecoup1} and \ref{eq:2HDMIdecoup2}), the Type-II
2HDM (Eq.~\ref{eq:2HDMIIdecoup}), the flipped and lepton-specific
2HDMs, and the democratic 3HDM.  In all models except the SM+S we set
$\tan\beta = 5$; for the democratic 3HDM we also set $\sin\Omega =
0.2$ (corresponding to $v_{\ell} = 50$ GeV) and $\cos\gamma = 0.5$.

On the right-hand side of each plot in Fig.~\ref{fig:hmodels-delta} we
also show the expected $1\sigma$ measurement uncertainties of the
squared Higgs couplings at the LHC from
Refs.~\cite{Duhrssen:2004uu,Duhrssen:2004cv} (summarized in
Table~\ref{tab:exptuncerts}), assuming SM coupling strengths and a
Higgs mass of 120~GeV.  The coupling fit of
Refs.~\cite{Duhrssen:2004uu,Duhrssen:2004cv} was based on anticipated
Higgs production and decay rate measurements from the LHC using 300
fb$^{-1}$ of integrated luminosity times two detectors.  Vector boson
fusion channels (which have only been studied for 30 fb$^{-1}$ to
date) are scaled to 100 fb$^{-1}$ to account for potential degradation
at high luminosity running.  The fit assumes SM rates in all channels
and allows for an unobserved component of the Higgs total width as
well as nonstandard contributions to the $ggh$ and $h \gamma\gamma$
vertices.  It further assumes $\bar g_W^2 = \bar g_Z^2 \leq 1.05$,
which is valid for models containing only doublets and/or singlets.
Theoretical uncertainties on Higgs production rates due to QCD scale
uncertainty were also included.  

\begin{table}
\begin{tabular}{lcccc}
\hline \hline
& $g_W^2$ & $g_b^2$ & $g_t^2$ & $g_{\tau}^2$ \\
\hline
LHC~\cite{Duhrssen:2004cv} & 22\% & 43\% & 32\% & 27\% \\
ILC~\cite{Djouadi:2005gi} & 2.4\% & 4.4\% & 6.0\% & 6.6\% \\
\hline \hline
\end{tabular}
\caption{\label{tab:exptuncerts}
Expected uncertainties on Higgs coupling-squared measurements
at the LHC and ILC, assuming $m_h = 120$~GeV and SM rates for all
processes involved.  See text for details.}
\end{table}

The results of Refs.~\cite{Duhrssen:2004uu,Duhrssen:2004cv} will
likely change when updated experimental and theoretical results are
included.  Updates of all experimental channels are now available in
the CMS Physics TDR~\cite{Ball:2007zza} and the ATLAS Computing System
Commissioning (CSC) Notes~\cite{Aad:2009wy}.  In particular, the
critical $tth$, $h \to b \bar b$ channel is dead; more work is needed
on the experimental side to evaluate the potential of newly-proposed
$bb$ final state channels like $Wh$, $h \to b \bar
b$~\cite{Butterworth:2008iy}.  Progress has also been made on the higher-order corrections to the $gg \to h$ cross section~\cite{DawsonSeattle09}.  A new fit involving these more sophisticated results will require some
work.

\begin{figure}[htbp]
\begin{center}
\includegraphics[width=0.78\textwidth,angle=0]{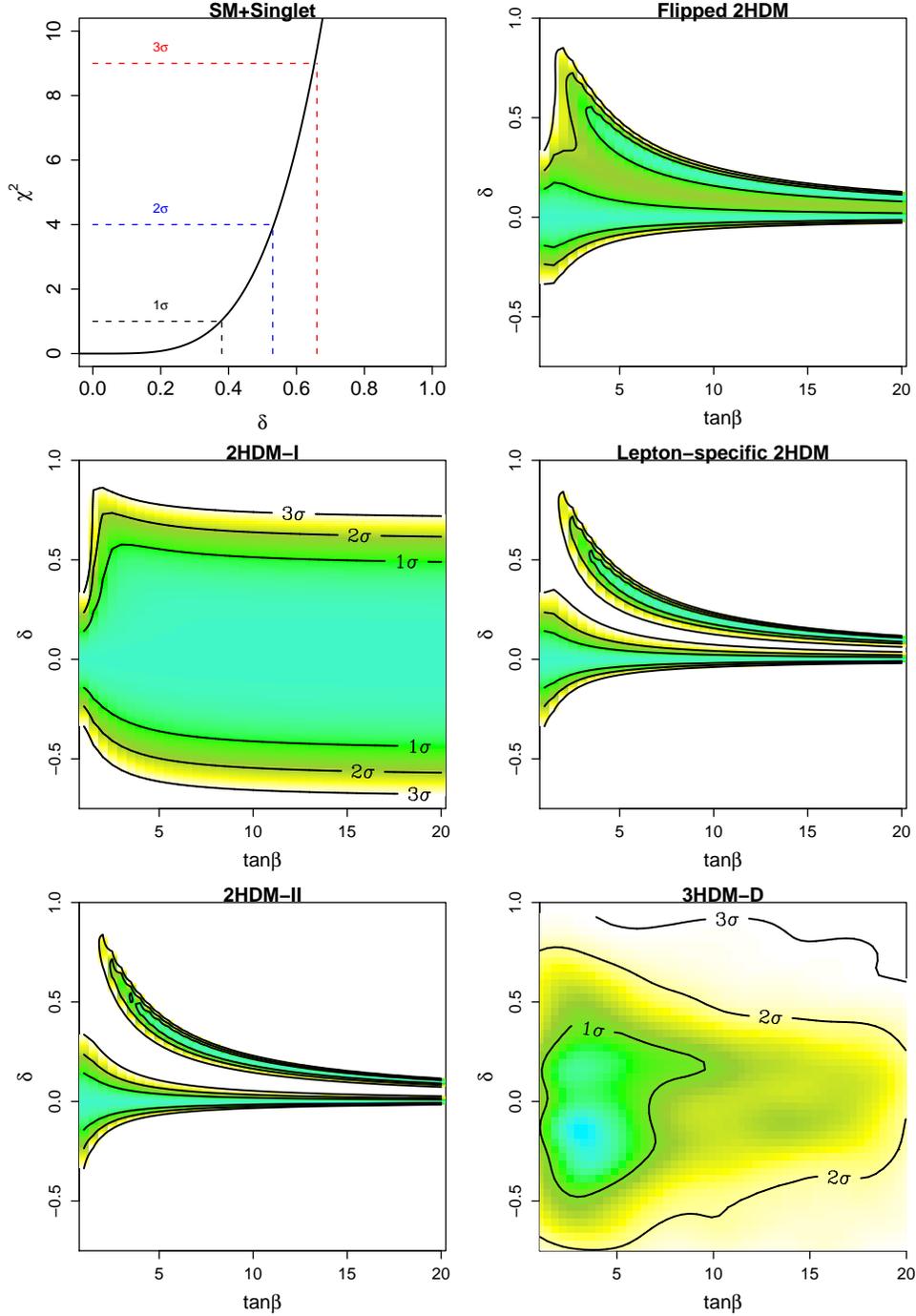}
\caption{Regions of parameter space with combined $h$ couplings within
one, two and three $\sigma$ corresponding to the inner, middle and
outer contours, respectively, of the SM limit for various models,
based on the expected LHC sensitivities given in
Table~\ref{tab:exptuncerts}.  Values of $\chi^2$ are calculated
according to Eq.~\ref{eq:chisq}.  The 3HDM-D model contains four free
parameters, $\delta$, $\tan\beta$, $\Omega$, and $\gamma$; we
marginalize over $\Omega$ and $\gamma$ by evaluating a Markov Chain
Monte Carlo (MCMC). The minor wiggles in the shapes of the contours
is due to the numerical precision of the MCMC.}
\label{fig:hmodels-decoup}
\end{center}
\end{figure}

\begin{figure}[h]
\begin{center}
\includegraphics[width=0.78\textwidth,angle=0]{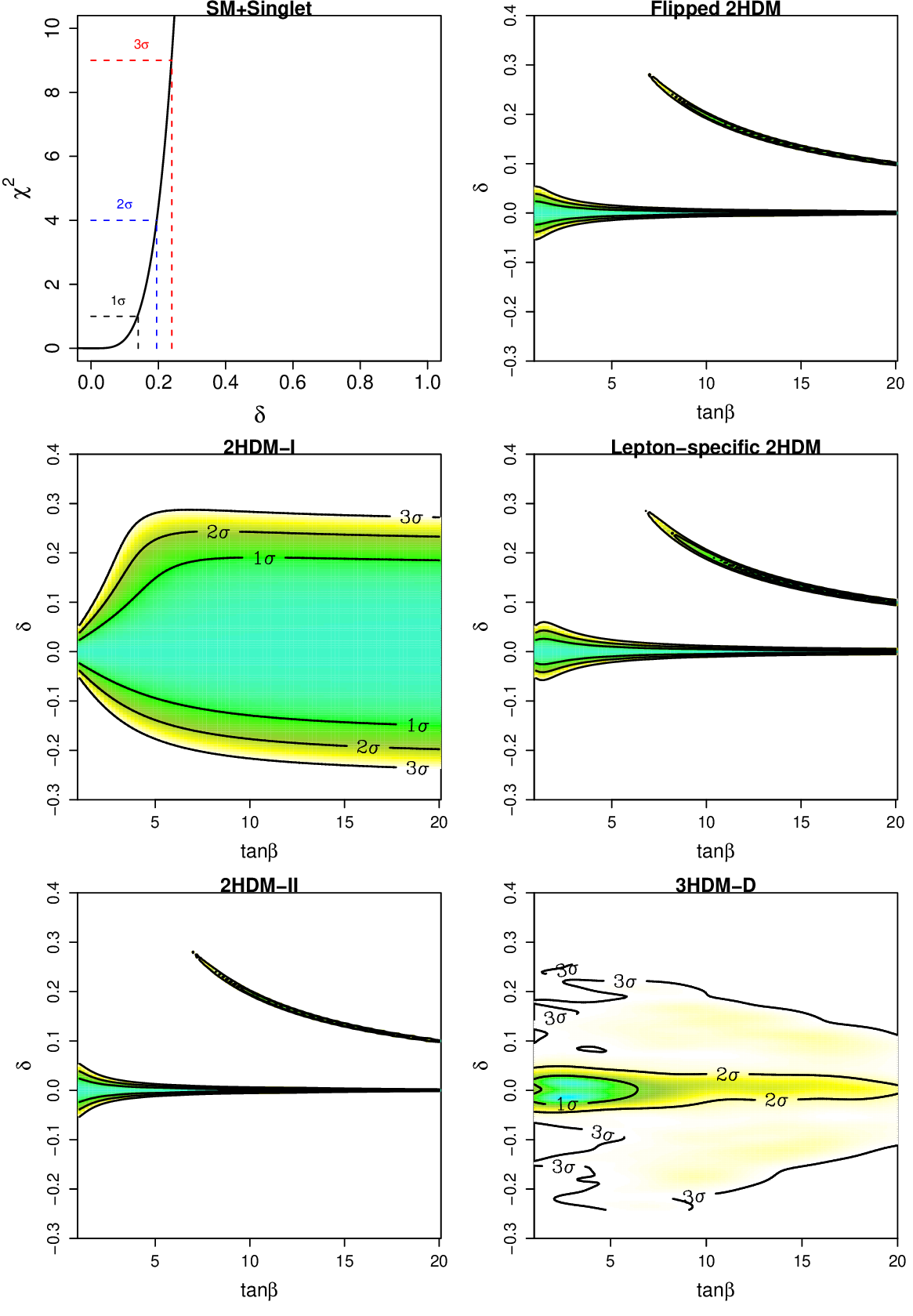}
\caption{Same as Fig.~\ref{fig:hmodels-decoup} but for the ILC, using
the precisions on couplings squared given in
Table~\ref{tab:exptuncerts}.}
\label{fig:hmodels-decoup-ilc}
\end{center}
\end{figure}

Nevertheless, we sketch the current situation as follows.  We scan
over model parameters and compute a $\chi^2$ relative to the SM
prediction according to
\begin{equation}
  \chi^2 = \sum_{i=W,b,t,\tau} \frac{(\Gamma_{i} - \Gamma_{i}^{SM})^2}
  {\left[ \delta \Gamma_i^{SM} \right]^2},
  \label{eq:chisq}
\end{equation}
using the LHC uncertainties in the partial widths from
Refs.~\cite{Duhrssen:2004uu,Duhrssen:2004cv} as summarized in
Table~\ref{tab:exptuncerts}.  We make no attempt here to account for
the correlations in the extracted couplings.  One-, two- and
three-sigma contours are shown in Fig.~\ref{fig:hmodels-decoup} for
the LHC.  For comparison, we show the corresponding ILC expectations
in Fig.~\ref{fig:hmodels-decoup-ilc}.

The SM plus a singlet contains only one additional parameter that
universally shifts the partial widths to all SM decay modes, while the
2HDM models listed contain $\delta$ and $\tan \beta$ as free
parameters that describe the Higgs coupling.  Since the 3HDM-D model
has four free parameters, $\delta$, $\tan\beta$, $\Omega$, and
$\gamma$, we marginalize over $\Omega$ and $\gamma$ by evaluating a
Markov Chain Monte Carlo
(MCMC)~\cite{Baltz:2006fm,Roszkowski:2007fd,Barger:2008qd} following
the procedure of Ref.~\cite{Barger:2008qd}.  

In summary, we have provided a first roadmap to determine the
underlying model of electroweak symmetry breaking under the assumption
that only Higgs doublets and/or singlets participate and the
Glashow-Weinberg-Paschos condition for natural flavor conservation
holds.  Our approach is based on the couplings of a single identified
CP-even Higgs state without regard to other Higgs particles that may
appear in the spectrum.  We restrict our considerations to tree-level
decays of the Higgs boson to avoid complications from new physics that
may appear in loop-mediated decays.  We described 15 classes of models
and compared their predictions for the shifts in the Higgs couplings
relative to the SM.  In each case, we presented formulae for the
couplings of a single CP-even state $h$ to $W$ or $Z$ boson pairs,
up-type quarks, down-type quarks, and charged leptons as a function of
the model parameters at tree level.  Where possible, we also inverted
those relations to provide explicit formulae for the model parameters
in terms of the $h$ couplings.  We summarized our results in a
decision tree that can be used to differentiate among the models.
While extraction of the couplings of $h$ with sufficient precision at
the LHC will be challenging, our results provide a starting point for
a more detailed study of model discrimination based on future
experimental results.


\begin{acknowledgments}
HEL and GS thank the organizers of the Brookhaven Forum 2007 for
providing a stimulating environment where this project was started.
VB was supported in part by the U.S.~Department of Energy under
grant No. DE-FG02-95ER40896 and by the Wisconsin Alumni Research Foundation.  GS was supported in party by the U.S.~Department of Energy under
grant Nos. DE-FG02-95ER40896  and DE-AC02-06CH11357.
HEL was supported by the Natural Sciences and Engineering Research
Council of Canada.
\end{acknowledgments}

\bibliographystyle{gsbib}
\bibliography{hmodels.bib}                      


\end{document}